\documentclass[aps,prl,twocolumn,superscriptaddress]{revtex4-2}
\pdfoutput=1
\usepackage{graphicx,amssymb,amsmath,amsfonts,empheq}
\usepackage[dvipsnames]{xcolor}
\usepackage{color}
\usepackage{braket}
\usepackage{latexsym}
\usepackage{upgreek}
\usepackage{dsfont}
\usepackage{bm}
\usepackage{makecell}
\usepackage{multirow}
\usepackage{enumitem}
\usepackage[normalem]{ulem}
\usepackage{url}
\usepackage{hyperref}
\hypersetup{
colorlinks = true,
linkcolor = [rgb]{0,0,0}, 
citecolor = [rgb]{0.25, 0.41, 0.88},
urlcolor = [rgb]{0.25, 0.41, 0.88}}
\usepackage{bbm}
\usepackage[utf8]{inputenc}
\usepackage[english]{babel}
\usepackage[T1]{fontenc}
\usepackage{lipsum}
\usepackage{physics}
\usepackage{tikz}
\usepackage{tikz-cd}
\usetikzlibrary{arrows}
\usetikzlibrary{intersections}
\usetikzlibrary{shapes.geometric}
\usetikzlibrary{decorations.pathmorphing, patterns,shapes}
\usetikzlibrary{decorations.markings}

\tikzset{
	partial ellipse/.style args={#1:#2:#3}{
		insert path={+ (#1:#3) arc (#1:#2:#3)}
	}
}

\tikzset{
	mid arrow/.style={postaction={decorate,decoration={
				markings,
				mark=at position .575 with {\arrow[#1]{stealth}}
	}}},
	near arrow/.style={postaction={decorate,decoration={
				markings,
				mark=at position .275 with {\arrow[#1]{stealth}}
	}}},
	far arrow/.style={postaction={decorate,decoration={
				markings,
				mark=at position .800 with {\arrow[#1]{stealth}}
	}}},
}
\pgfmathsetmacro\MathAxis{height("$\vcenter{}$")}

\pgfdeclarepatternformonly{south west lines}{\pgfqpoint{-0pt}{-0pt}}{\pgfqpoint{3pt}{3pt}}{\pgfqpoint{3pt}{3pt}}{
	\pgfsetlinewidth{0.4pt}
	\pgfpathmoveto{\pgfqpoint{0pt}{0pt}}
	\pgfpathlineto{\pgfqpoint{3pt}{3pt}}
	\pgfpathmoveto{\pgfqpoint{2.8pt}{-.2pt}}
	\pgfpathlineto{\pgfqpoint{3.2pt}{.2pt}}
	\pgfpathmoveto{\pgfqpoint{-.2pt}{2.8pt}}
	\pgfpathlineto{\pgfqpoint{.2pt}{3.2pt}}
	\pgfusepath{stroke}}

\newcommand{\ri}{\mathrm{i}}

\newcommand{\bbZ}{\mathbb{Z}}

\newcommand{\calF}{\mathcal{F}}

\newcommand{\calR}{\mathcal{R}}

\newcommand{\calZ}{\mathcal{Z}}

\newcommand{\sfa}{\mathsf{a}}

\definecolor{red1}{RGB}{240,83,90}
\definecolor{blue1}{RGB}{91,98,165}
\definecolor{myturquoise}{RGB}{83,195,189}
\definecolor{lightturquoise}{RGB}{64, 224, 208}
\definecolor{violet}{RGB}{207, 159, 255}
\definecolor{darkturquoise}{RGB}{54, 194, 178}  %
\definecolor{darkviolet}{RGB}{177, 129, 225}    %
\definecolor{canaryyellow}{RGB}{255, 255, 143}
\definecolor{mygreen}{RGB}{134, 175, 142}
\definecolor{mint}{RGB}{152, 251, 152}
\definecolor{mint2}{RGB}{218, 247, 166}
\definecolor{myred}{RGB}{240,83,90}
\definecolor{myblue}{RGB}{73,103,189}

\begin{document}

\title{Phases of decodability in the surface code with unitary errors}

\author{Yimu Bao}
\affiliation{Kavli Institute for Theoretical Physics, University of California, Santa Barbara, CA 93106, USA}

\author{Sajant Anand}
\affiliation{Department of Physics, University of California, Berkeley, CA 94720, USA}

\begin{abstract}
The maximum likelihood (ML) decoder in the two-dimensional surface code with generic unitary errors is governed by a statistical mechanics model with complex weights, which can be simulated via (1+1)D transfer matrix contraction. Information loss with an increasing error rate manifests as a ferromagnetic-to-paramagnetic transition in the contraction dynamics. In this work, we establish entanglement as a \textit{separate} obstruction to decoding; it can undergo a transition from area- to volume-law scaling in the transfer matrix contraction with increasing unitary error rate. In particular, the volume-law entanglement can \textit{coexist} with ferromagnetic order, giving rise to a phase in which the encoded information is retained yet is effectively undecodable. We numerically simulate the ML decoding in the surface code subject to both single- and two-qubit Pauli-X rotations and obtain a phase diagram that contains a ferromagnetic area-law, a paramagnetic volume-law, and a potential ferromagnetic volume-law phase. We further show that, starting from the paramagnetic volume-law phase, tilting the single-qubit rotation away from the X-axis couples the stat-mech models for X and Z errors and can lead to a ferromagnetic volume-law phase in which, although Z errors remain in principle correctable, the encoded classical information is hard to recover. To perform numerical simulations, we develop an algorithm for syndrome sampling based on an isometric tensor network representation of the surface code.
\end{abstract}

\maketitle

\emph{Introduction.}---
The surface code is a promising candidate for quantum memory~\cite{kitaev2003fault,dennis2002topological} and has been realized in several leading experimental platforms~\cite{semeghini2021probing,bluvstein2022quantum,satzinger2021realizing,google2023suppressing,andersen2023observation,bluvstein2023logical}.
One of its striking properties as a topological code is the robust encoding of quantum information up to a finite threshold of local errors. 
In a seminal work, Dennis et al. showed that in the two-dimensional surface code with \textit{incoherent} errors, the finite error threshold achieved by the optimal decoding algorithm maps to the transition point in the two-dimensional random bond Ising model (RBIM) on the Nishimori line~\cite{dennis2002topological,nishimori1981internal}.
The optimal decoding algorithm, known as the maximum likelihood (ML) decoder, evaluates the probability of each class of homologically equivalent error strings compatible with the observed error syndromes and then applies a recovery operator to correct the most probable error class~\footnote{The ML decoding we consider is sometimes referred to in literature as \textit{degenerate} quantum ML (DQML) decoding~\cite{iyer2015hardness,deMarti_iOlius_2024}. Degeneracy is not found in classical codes and indicates that multiple error strings can give rise to the same error syndrome.}.
The total probability for each homological class corresponds to the partition function of an RBIM, which undergoes a ferromagnetic-to-paramagnetic (FM-PM) transition that leads to a sharp change of decoding fidelity at the error threshold~\cite{terhal2015quantum}.

\begin{figure}[t!]
\centering
\begin{tikzpicture}[scale=0.7]
\foreach \y in {0,1,...,3}{
    \draw[black!30,line width=1.] (-5.7,\y+0.5) -- (-1.1,\y+0.5);
}
\foreach \x in {0.5,1.5,...,3.5}{
    \draw[black!30,line width=1.] (\x-5.4,0.5) -- (\x-5.4,3.5);
}
       
\foreach \y in {0.5,1.5,...,3}{
\foreach \x in {0.5,1.5,...,3.5}{
\draw[black!30, thick, fill=black!20] (\x-5.4,\y+0.5) circle (0.05);
}
}
\foreach \y in {0,1,...,3}{
\foreach \x in {0,1,...,4}{
\draw[black!30, thick, fill=black!20] (\x-5.4,\y+0.5) circle (0.05);
}
}

\draw[myred,line width=1.5] (-4.4,2.5) -- (-3.4,2.5);
\draw[myred,line width=1.5] (-3.9,2) -- (-3.9,3);
\node[myred] at (1.8-5.4,2.8) {$A_v$};
\filldraw[myred] (-4.4,2.5) circle (0.06);
\filldraw[myred] (-3.4,2.5) circle (0.06);
\filldraw[myred] (-3.9,2) circle (0.06);
\filldraw[myred] (-3.9,3) circle (0.06);

\draw[myblue,line width=1.5] (2.5-5.4,3+0.5) -- (3.5-5.4,3+0.5) -- (3.5-5.4,2+0.5) -- (2.5-5.4,2+0.5) -- cycle;
\node[myblue] at (-2.4,3) {$B_p$};
\filldraw[myblue] (-2.4,3.5) circle (0.06);
\filldraw[myblue] (-1.9,3) circle (0.06);
\filldraw[myblue] (-2.4,2.5) circle (0.06);
\filldraw[myblue] (-2.9,3) circle (0.06);

\draw[darkviolet,line width=1.5] (0.7-5.4,0.5) -- (1.5-5.4,0.5) node[below,darkviolet]{$e^{\ri\varphi X_\ell X_{\ell'}}$} -- (2.3-5.4,0.5);
\filldraw[darkviolet] (1-5.4,0.5) circle (0.06);
\filldraw[darkviolet] (2-5.4,0.5) circle (0.06);

\draw[darkturquoise,line width=1.5] (3.5-5.4,0.7) -- (3.5-5.4,1.0) node[right,darkturquoise]{$e^{\ri\theta X_\ell}$} -- (3.5-5.4,0.8+0.5);
\filldraw[darkturquoise] (3.5-5.4, 1) circle (0.06);
\draw[darkturquoise,line width=1.5] (-2.7,1.5) -- (-2.1,1.5);
\filldraw[darkturquoise] (-2.4, 1.5) circle (0.06);
\node at (-6.1,3.5) {(a)};
\foreach \x in {1,...,3,4}{
\draw[blue1!30,line width=1] (\x, 0.5) -- (\x, 3.5);
}
\foreach \y in {0.5,1.5,2.5,3.5}{
\draw[blue1!30,line width=1] (0,\y) -- (4.9,\y);
}
\foreach \x in {0.5,1.5,2.5,3.5,4.5}{
\foreach \y in {0.5,1.5,2.5,3.5}{
\filldraw[rounded corners=1pt, draw=black!80, fill=lightturquoise] (\x-0.1,\y-0.1) rectangle ++ (0.2,0.2);
}
}
\foreach \x in {1,2,3,4}{
\foreach \y in {1,2,3}{
\filldraw[rounded corners=1pt, draw=black!80, fill=lightturquoise] (\x-0.1,\y-0.1) rectangle ++ (0.2,0.2);
}
}
\foreach \x in {1,...,4}{
\foreach \y in {0.5,1.5,2.5,3.5}{
\filldraw[blue1] (\x,\y) circle (0.05);
\draw[violet, line width=1pt] (\x+0.25,\y) arc[start angle=-30, end angle=-150, radius=0.28867513459];
\filldraw[rounded corners=1pt, draw=black!80, fill=violet] (\x-0.25-0.075,\y-0.075) rectangle ++ (0.15,0.15);
\filldraw[rounded corners=1pt, draw=black!80, fill=violet] (\x+0.25-0.075,\y-0.075) rectangle ++ (0.15,0.15);
}
}
\foreach \x in {-0.1,5}{
\foreach \y in {0.5,1.5,2.5,3.5}{
\draw[-latex, thick, blue1] (\x-0.1,\y-0.1) -- ++(0.3,0.3);
}
}
\node[black] at (-0.6,3.5) {(b)};
\node at (-2.8,-3.8) {\includegraphics[width=2in]{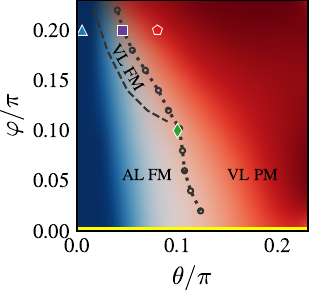}};
\node[black] at (-6.1,-0.5) {(c)};
\node at (3.3,-3.8) {\includegraphics[width=1.2in]{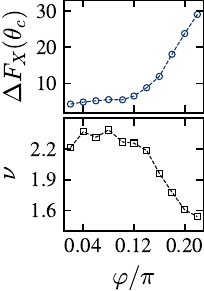}};
\node[black] at (1.4,-0.5) {(d)};
\end{tikzpicture}
\caption{(a) Two-dimensional surface code on the square lattice with rough and smooth boundary conditions in $x$ and $y$ direction, respectively. Qubits are placed on the edges and are subject to single-qubit $e^{\ri\theta X_\ell}$ and horizontal two-qubit Pauli-X rotations $e^{\ri\varphi X_\ell X_\ell'}$ on primal lattice rows. (b) Tensor network representation of the partition function of the random bond Ising model for maximum likelihood decoding. 
(c) Phase diagram for ($\varphi > 0$). The ferromagnetic (FM) and paramagnetic (PM) phases are the encoding and information-loss regimes. The volume-law (VL) and area-law (AL) entanglement scaling characterize the hardness of decoding.
On the x-axis (highlighted in yellow), the decoding is governed by a free-fermion network model, leading to qualitatively different physics.
(d) Defect free energy and correlation length critical exponents on the critical line $(\theta_c,\varphi_c)$ as a function of $\varphi$.
}
\label{fig:1}
\end{figure}

Errors in realistic quantum devices are generally \textit{coherent}, i.e. creating coherence between syndrome configurations, due to, e.g., imperfect unitary gates or tilted measurement basis in the state preparation~\cite{bravyi2018correcting}.
Coherent errors lead to a key difference in ML decoding as the probability for each class is now a sum over generally \textit{complex} Boltzmann weights.
This prevents identifying the most likely class using efficient Monte Carlo algorithms.
Instead, one may evaluate the complex partition function using the transfer matrix method, also known as tensor network decoding~\cite{ferris2014tensor,bravyi2014efficient,Darmawan_2017,Darmawan_2018,tuckett2018ultrahigh,tuckett2019tailoring,chubb2021statistical,chubb2021general,darmawan2024optimaladaptationsurfacecodedecoders}.

Recently, Ref.~\cite{behrends2022surface} provided an alternative perspective, formulating the decoding transition as a ferromagnetic transition in (1+1)D non-unitary dynamics generated by the transfer matrix.
This formulation raises the possibility of a separate entanglement transition similar to the MIPT in the monitored dynamics, where the bipartite entanglement in trajectories can undergo a phase transition from a volume- to an area-law scaling tuned by the measurement rate~\cite{li2018quantum,skinner2019measurement,li2019measurement,choi2020quantum,gullans2020dynamical,potter2022entanglement,fisher2023random,sang2021measurement,alberton2021entanglement}.
Such a transition would in general be associated to the hardness of performing ML decoding by simulating the transfer matrix dynamics.
However, the (1+1)D dynamics associated with the specific error models considered in Ref.~\cite{behrends2022surface} are Gaussian fermion dynamics, which are non-generic and efficient to simulate on classical computers.
It is natural to ask, for generic coherent errors, whether the transfer matrix dynamics can exhibit a phase with volume-law entanglement and become hard to simulate.

In this Letter, we focus on the surface code with coherent errors that are purely unitary rotations and demonstrate that general unitary errors can lead to volume-law entanglement in the transfer matrix contraction, preventing efficient ML decoding.
In particular, the volume-law entanglement can coexist with the ferromagnetic order in RBIM, giving rise to a phase in which the information is in principle retained but practically undecodable.
We identify such a possibility in two different cases.

First, we consider the unitary errors that consist of single- and two-qubit Pauli-X rotations.
We show that large rotation angle errors can result in volume-law entanglement in the transfer matrix contraction.
The phase diagram contains (1) a ferromagnetic, encoding, area-law phase, (2) a paramagnetic, corrupted, volume-law phase, and (3) a potential ferromagnetic, encoding, volume-law phase as shown in Fig.~\ref{fig:1}(c).

Next, we start from the paramagnetic volume-law phase in Fig.~\ref{fig:1}(c) and tilt the single-qubit rotation into the XY plane.
For the rotation axis sufficiently close to the X-axis, the surface code is a classical memory as the Pauli-Z errors remain correctable.
However, the coupling between two RBIMs for correcting Pauli-Z and X errors gives rise to a ferromagnetic volume-law phase, as recovering the classical information now requires simulating the transfer matrix contraction with volume-law entanglement.

To numerically simulate ML decoding for generic unitary errors, we develop an algorithm that allows sampling syndromes from a corrupted surface code.
Our algorithm is based on an isometric tensor network representation of the surface code and converts syndrome sampling into simulating (1+1)D monitored dynamics~\cite{zaletel2020isometric}.

\emph{Surface code and error model.}--- 
We consider the 2D surface code on the square lattice as shown in Fig.~\ref{fig:1}(a).
The surface code involves two types of stabilizers $A_v = \prod_{\ell \in \text{star}(v)}X_\ell$ and $B_p = \prod_{\ell \in \partial p} Z_\ell$ associated with the vertex $v$ and the plaquette $p$.
We take the smooth and rough boundary conditions in $x$ and $y$ direction, respectively, and include incomplete, three-body star and plaquette boundary operators in the stabilizer group.
The logical space is the simultaneous $+1$ eigenspace of the stabilizers, which is twofold degenerate and encodes one qubit of quantum information.
The logical X (Z) operator is given by the product of Pauli-X (Z) along the vertical (horizontal) direction on the dual (primal) lattice. 

We consider an error model of arbitrary single-qubit unitary rotations and two-qubit Pauli-X rotations, giving rise to a rotated state 
\begin{align}
    \ket{\Psi} = \prod_{\langle \ell,\ell'\rangle_h} e^{\ri\varphi X_\ell X_{\ell'}} \prod_\ell e^{\ri \theta \mathbf{n}\cdot \mathbf{S}_\ell}\ket{\Psi_0}, 
    \label{eq:error_model}
\end{align}
with rotation angles $\theta, \varphi \in [0, \pi/4]$, rotation unit vector $\mathbf{n} = (n_x, n_y, n_z)$, and $\mathbf{n}\cdot\mathbf{S}_\ell = n_x X_\ell+n_y Y_\ell + n_z Z_\ell$.
For simplicity, we consider two-qubit Pauli-X rotations acting only on neighboring qubits on horizontal edges [Fig.~\ref{fig:1}(a)] and assume the rotation angle $\theta$ and $\varphi$ are spatially uniform~\footnote{As discussed later, horizontal two-qubit (2Q) noise is sufficient to break the Gaussianity of the single-qubit X-type Pauli rotations. As (1) horizontal two-qubit noise is simpler to implement in our isometric tensor network sampling than vertical or nearest-neighbor 2Q errors and (2) we do not expect our results to qualitatively depend on the choice of non-Gaussianity, we restrict to horizontal 2Q errors.}.

\emph{Maximum likelihood decoding.---}
Error correction in the surface code starts with the stabilizer measurements, which have random outcomes according to the Born rule.
The stabilizer measurements, called the \textit{syndrome}, with $-1$ outcomes reveal information about the error operator.
Based on the acquired syndrome, the decoding algorithm infers the probable X (and Z) error string, which falls into two homological classes that differ by a logical-X (Z) operator along a non-contractible path.

Considering only $X$-type errors for now, the ML decoder first computes the total probability $P_{\mathsf{a},s}$ of all error strings in each of two classes $\mathsf{a} = 0,1$ for the given syndrome $s$.
Then, one applies a recovery operator from the most likely class to restore the encoded state.
Such an algorithm has an average decoding fidelity
\begin{align}
    \calF = \sum_s P_s \frac{\max\{P_{0,s}, P_{1,s}\}}{P_{0,s} + P_{1,s}},
\end{align}
where the probability of syndrome $s$ is $P_s = P_{0,s} + P_{1,s}$.

The probability of each class maps to the partition function of a 2D RBIM~\cite{dennis2002topological}.
The mapping has been generalized to various topological codes subject to local incoherent errors~\cite{wang2003confinement,katzgraber2009error,bombin2012strong,kubica2018three,chubb2021statistical,song2022optimal}.
Specifically, one first chooses a string $\calR_{\mathsf{a},s}$ from the decoding class $\mathsf{a}$, which matches the observed syndrome $s$, i.e. $\partial \calR_{\mathsf{a},s} = s$.
The partition function $\calZ_{\mathsf{a},s}$ with negative bonds $\eta_{ij} = -1$ along the string $\calR_{\mathsf{a},s}$ gives the unnormalized total probability $P_{\sfa,s}$ of error strings in class $\mathsf{a}$.
We note that, although the choice of $\calR_{\mathsf{a},s}$ has a gauge redundancy, $\calZ_{\sfa,s}$ is a class function taking the same value for homologically equivalent $\calR_{\sfa,s}$.

The decoding fidelity depends on the ratio between the partition functions for two classes, which is given by the excess free energy of defect insertion along a homologically nontrivial path, i.e. $\Delta F = \sum_s P_s |\log(\calZ_{1,s}/\calZ_{0,s})|$.
When increasing the error rate, the RBIM undergoes a FM-PM transition detected by the defect free energy, which scales linearly with the system size in the ferromagnetic phase and takes a vanishing value in the paramagnetic phase.
This indicates a decoding transition as the decoding fidelity sharply changes from unity to $1/2$.
Importantly, the defect free energy is scale-free at a critical point with conformal symmetry and can be used to accurately determine the error threshold~\cite{honecker2001universality}.

\begin{figure*}[t!]
    \centering
    \includegraphics[width=\textwidth]{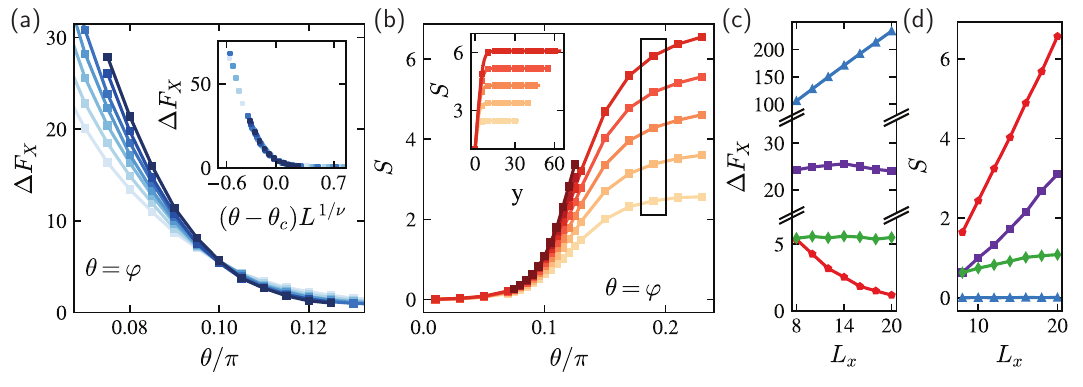}
    \caption{Decoding transition in the surface code under single- and two-qubit Pauli-X rotations. (a) Defect free energy as a function of $\theta$ along the cut $\varphi = \theta$ in the phase diagram. (a, inset) Finite-size scaling using the ansatz $\Delta F_X = f((\theta - \theta_c)L^{1/\nu})$ with $\theta_c = 0.102\pi$ and $\nu = 2.27$. (b) Half-system entanglement entropy in transfer matrix contraction as a function of $\theta$ for $\varphi = \theta$. (b, inset) Entanglement as a function of contraction steps $y$ for $\theta = \varphi = 0.19\pi$. Curves with increasing opacity label system sizes $L = 8,10,12, \cdots,20$ with $L_y = 4L_x = 4L$. (c, d) Defect free energy (panel c) and entanglement entropy (panel d) as a function of system size $L_x$ with $L_y = 4L_x$ for $(\theta,\varphi) = (0.1\pi,0.1\pi)$ (green diamond), $(0.005\pi, 0.2\pi)$ (blue triangle), $(0.045\pi, 0.2\pi)$ (purple square), $(0.08\pi, 0.2\pi)$ (red pentagon) marked in Fig.~\ref{fig:1}(c). The bond dimension is up to $\chi = 512$ in the simulation. The results are averaged over up to $1000$ syndrome realizations, and statistical error bars are within the marker size.
    }
    \label{fig:2}
\end{figure*}

ML decoding for coherent errors has a key difference from that for incoherent errors; the probability $P_{\mathsf{a},s}$ of error strings in each class is given by a partition function with generally \textit{complex} Boltzmann weights~\cite{behrends2022surface,venn2022coherent}.
Evaluating the RBIM with complex weights can no longer be done with Monte Carlo methods and instead relies on the transfer matrix method, which converts the problem to that of simulating (1+1)D quantum dynamics generally involving non-unitary projectors and unitary gates~\cite{ferris2014tensor,bravyi2014efficient,tuckett2018ultrahigh,tuckett2019tailoring,chubb2021statistical,chubb2021general,behrends2022surface}.

The computational cost of simulating transfer matrix contraction scales exponentially with the entanglement in the (1+1)D dynamics~\cite{verstraete2006matrix,schuch2008entropy},
\begin{align}
    S = \sum_s P_s S_s,
\end{align}
where $S_s$ is the half-system entanglement entropy for syndrome $s$.
We remark that although the transfer matrix in the RBIM depends on the gauge choice, i.e. $\calR_{\sfa,s}$, of recovery string guess and thus random bond configuration, the entanglement in the contraction is gauge invariant~\cite{SOM}.

\emph{Numerical results.}---
We consider the surface code containing $L_x \times L_y$ plaquettes and subject to the unitary errors in Eq.~\eqref{eq:error_model}.
We sample syndromes from the corrupted surface code, derive the statistical mechanics model for ML decoding, and evaluate its partition function using the transfer matrix method~\cite{SOM}.
In numerical simulation, we represent the intermediate 1D state during the transfer matrix contraction as a matrix product state and perform the contraction using the time-evolving block-decimation (TEBD) algorithm~\cite{vidal2004efficient}. 

The simplest unitary error in the surface code is the single-qubit Pauli-X rotation~\cite{bravyi2014efficient,Darmawan_2017,venn2022coherent,behrends2022surface}.
In this case, the transfer matrix contraction maps to a parity-conserving Gaussian fermion dynamics that involves unitary gates and projections~\cite{bravyi2018correcting,behrends2022surface}.
Such dynamics can be efficiently simulated and also do not exhibit a volume-law entanglement~\cite{cao2019entanglement,alberton2021entanglement,bao2021symmetry,jian2022criticality,fava2023nonlinear,jian2023measurement}.
However, these features are non-generic; tilting the rotation axis away from Pauli-X or introducing two-qubit Pauli-X rotations breaks Gaussianity and, in principle, can lead to a volume-law phase.

To begin, we consider the surface code under single- and two-qubit Pauli-X rotations with rotation angles $\theta$ and $\varphi$, respectively [i.e. $\mathbf{n} = (1,0,0)$ in Eq.~\eqref{eq:error_model}].
The RBIM for correcting X errors consists of Ising spins on the vertices of the square lattice and involves coupling between neighboring Ising spins and additional next-nearest-neighbor coupling in the horizontal direction [illustrated in Fig.~\ref{fig:1}(b)]~\cite{SOM}.
The ferromagnetic transition in RBIM can be accurately determined using the defect free energy, which is scale-free at the critical point: 
Along the cut $\theta = \varphi$, the numerically obtained defect free energies for various system sizes cross at a single point $\theta_c= \varphi_c = 0.102\pi$ as shown in Fig.~\ref{fig:2}(a).
Finite-size scaling analysis with ansatz $\Delta F_X = f((\theta - \theta_c)L^{1/\nu})$ yields a scaling collapse with $\nu = 2.27$ [Fig.~\ref{fig:2}(a, inset)].
We further take various horizontal cuts in $\theta$ (with fixed $\varphi$) of the phase diagram to determine the ferromagnetic phase boundary and observe transitions with the same university for $\varphi \lesssim 0.1\pi$.
Specifically, the correlation length exponent $\nu \simeq 2.3$ and the defect free energy $\Delta F_X(\theta_c,\varphi_c) \simeq 5$ remain approximately unchanged [Fig.~\ref{fig:1}(d)].
Interestingly, for $\varphi \gtrsim 0.1\pi$, the university of ferromagnetic transition becomes $\varphi$ dependent, indicating a possible multi-critical point at $\theta_c = \varphi_c = 0.102\pi$.

Half-system entanglement entropy exhibits distinct scaling on the two sides of the transition on the $\theta = \varphi$ cut [Fig.~\ref{fig:2}(b)], indicating a ferromagnetic, area-law phase for small $\theta$ and a paramagnetic, volume-law phase for large $\theta$.
However, determining the entanglement transition point solely based on the entanglement data is challenging due to the lack of scale-free probes in the numerics based on matrix product states, the challenge of probing volume-law entanglement at scale due to finite classical resources, and the difficulty in distinguishing area- vs. log- vs volume-law entanglement from small system sizes.
Instead, we examine the entanglement scaling with system size across the FM-PM transition to determine whether the entanglement and the FM-PM transition happen simultaneously.

Along the ferromagnetic phase boundary, we observe a logarithmic entanglement scaling at the critical point for $\varphi \lesssim 0.1\pi$, consistent with a direct transition between the ferromagnetic, area-law phase and paramagnetic, volume-law phase [Fig.~\ref{fig:2}(d)].
For larger $\varphi$, however, we find both volume-law entanglement scaling at the critical point [e.g., at $(\theta_c,\varphi_c) = (0.045\pi,0.2\pi)$ in Fig.~\ref{fig:2}(d)] and a change in the universality of FM-PM transition (Fig.~\ref{fig:1}(d)).
This indicates the potential ferromagnetic volume-law phase sketched in Fig.~\ref{fig:1}(c).
This volume-law entanglement prevents an efficient recovery of the in principle still encoded quantum information in the surface code.

The possibility of a ferromagnetic volume-law phase was pointed out in the context of MIPT in Ref.~\cite{bao2021symmetry}.
Moreover, since the $\bbZ_2$ ordering and entanglement transition are governed by the physical $\bbZ_2$ and the replica symmetry in the effective model for MIPT, respectively, these two transitions need not occur at the same point.
However, beyond numerical evidence, we do not have a compelling argument for the existence of such a phase in this specific model, and we cannot exclude the possibility of finite-size effects.

\begin{figure}
    \centering
    \includegraphics[width=\linewidth]{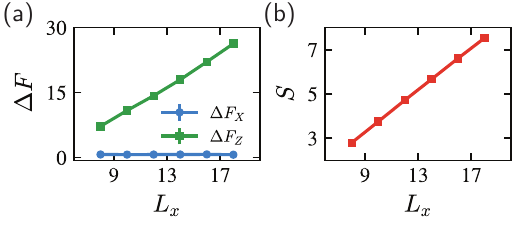}
    \caption{Decoding in the surface code under single-qubit unitary rotation $e^{\ri\theta_x X_\ell+\ri\theta_y Y_\ell}$ and two-qubit Pauli-X rotation $e^{\ri\varphi X_\ell X_{\ell'}}$. We choose a fixed $\theta_x = 0.2\pi$, $\theta_y = 0.06\pi$, and $\varphi = 0.2\pi$. (a) Excess free energies for inserting defects associated with logical $X$ and $Z$ operators as a function of system size $L$ with $L_y = 2L_x = 2L$. (b) Half-system entanglement entropy $S$ as a function of system sizes in the steady state.
    }
    \label{fig:3}
\end{figure}

Next, we present another mechanism leading to a ferromagnetic volume-law phase. 
We start with the surface code subject to single- and two-qubit X-rotations with $\theta_x = \varphi = 0.2\pi$, deep in the paramagnetic, volume-law phase.
We now tilt the rotation axis of the single-qubit error into the XY plane, s.t. $e^{\ri\theta\mathbf{n}\cdot\mathbf{S}_\ell} = e^{\ri\theta_x X_\ell + \ri\theta_y Y_\ell}$ as in Eq.~\eqref{eq:error_model}.
In this case, the RBIM involves two sets of Ising spins associated with X and Z error strings and therefore exhibits a $\bbZ_2 \times \bbZ_2$ symmetry.
When increasing $\theta_y$, the excess free energy $\Delta F_Z$ of a defect insertion in the Ising spins associated with Z errors detects a $\bbZ_2$ symmetry-breaking transition, indicating the loss of \textit{classical} information.
At $\theta_y = 0.06\pi$ which is below the threshold $\theta_{y,c} \simeq 0.13 \pi$, we observe both a linear scaling of the defect-free energy [Fig.~\ref{fig:3}(a)] associated with Z logical and volume-law entanglement [Fig.~\ref{fig:3}(b)]~\cite{SOM}.
This indicates that the classical information, although retained in the surface code (since Z errors are correctable), becomes practically hard to recover due to coupling to the X error RBIM in the volume-law phase.

The ferromagnetic, volume-law phase when tilting the single-qubit rotation axis is similar to that realized in the monitored random circuit with $\bbZ_2$ symmetry~\cite{bao2021symmetry,li2023decodable}.
The transfer matrix contraction for the ML decoder has a natural two-chain structure for correcting X and Z errors in the surface code.
Here, for single-qubit X-rotation ($\theta_y = 0$), the RBIMs for X and Z errors are decoupled and exhibit volume-law entanglement and $\bbZ_2$ symmetry breaking, respectively.
Tilting the rotation axis couples the two RBIMs together.
Since both volume-law and $\bbZ_2$ ordered phases have finite correlation length and are understood as \emph{gapped} phases in the effective model for the monitored circuit, one expects them to be stable under weak coupling between the two chains~\cite{bao2021symmetry}.

\emph{Discussion.}---
ML decoding in the surface code with general unitary errors relies on accurate transfer matrix contraction to evaluate the probability of each homological class.
In this work, we have demonstrated that the contraction, specifically using matrix product states, can be computationally hard due to the generated volume-law entanglement.
In two different cases, we numerically identified the coexistence of volume-law entanglement with ferromagnetic ordering, indicating that the encoded information is retained and in principle decodable, yet hard to efficiently recover.

Our work opens several directions for future studies.
First, our work demonstrates that the encoded information in the surface code with unitary errors can be irrecoverable in practice either fundamentally due to information loss or computationally due to large-scale entanglement in the transfer matrix contraction.
Although the encoded information, quantified by the coherent information, is monotonically non-increasing under post-processing~\cite{nielsen2010quantum}, the entanglement in the contraction is not a monotone. 
It is natural to ask whether one can reduce the entanglement and extend the practically decodable regime by e.g. deliberately introducing additional incoherent errors. 

Next, it is worth extending the discussion to the surface code with syndrome measurement errors. 
In this case, retaining information requires repeated syndrome measurements, and the probability for each decoding class is governed by a three-dimensional statistical mechanics model with possible complex weights or equivalently $(2+1)$-dimensional quantum dynamics~\cite{dennis2002topological}.
Here, the unitary evolution introduced by the coherent error can create hardness in classical simulation even if the dynamics only exhibit area-law entanglement.

Moreover, the transition in information recoverable from a corrupted surface code or general decohered topological order has been formulated as an intrinsic mixed-state transition~\cite{bao2023mixed,fan2023diagnostics,lee2023quantum}. 
This transition in the case of unitary errors corresponds to the loss of information in the ensemble of pure states after syndrome measurements. 
It is interesting to ask whether diagnostics intrinsic to the quantum state and syndrome measurement record can detect the entanglement transition in the contraction. Can such a transition, from a dual perspective, be determined by the distribution of syndrome outcomes? 

Finally, the complex-weight RBIM derived for the surface code under unitary errors features a generalized Nishimori condition~\cite{fan2023diagnostics} and presents novel critical points that can go beyond the conventional unitary conformal field theory. 
It is worth examining the conformal symmetry and determining the conformal data at the critical point using transfer matrix contraction based on the TEBD algorithm~\cite{pollmann2009theory}. 

\begin{acknowledgments}
\emph{Note}.---
Upon finishing the manuscript, we became aware of a related paper~\cite{behrends2024statistical} on arXiv studying the maximum likelihood decoding in the surface code under unitary errors. 
Our paper presents a different algorithm that samples the syndrome directly from the corrupted state, and demonstrates that purely unitary errors, in general, can lead to a phase in which decoding is hard.

We also note a concurrent work~\cite{cheng2025emergent}, which proposes an algorithm, similar to ours, to simulate the logical error when decoding the surface code with coherent errors and observes an entanglement transition in the simulation.

\emph{Acknowledgements}.--- 
We thank Ehud Altman, Zhehao Dai, Junkai Dong, Ruihua Fan, Matthew Fisher, Sam Garratt, Sarang Gopalakrishnan, Michael Gullans, Bryce Kobrin, Ali Lavasani, Yaodong Li, and Sagar Vijay for helpful discussions. 
We especially thank Matthew Fisher, Jacob Hauser, Yaodong Li, and Sagar Vijay for collaboration on related projects.
SA thanks Michael Zaletel for generous support and Michael Zaletel, Yantao Wu, and Sowmya Thanvantri for collaboration on a related project.
Y.B. is supported in part by grant NSF PHY-2309135 and the Gordon and Betty Moore Foundation Grant No. GBMF7392 to the Kavli Institute for Theoretical Physics (KITP).
S.A. is supported by the U.S. Department of Energy, Office of Science, Basic Energy Sciences, under Early Career Award No. DE-SC0022716.
This research was done using services provided by the OSG Consortium~\cite{osg1,osg2,osg3,osg4}, which is supported by the National Science Foundation awards No. 2030508 and No. 1836650.
Computing resources were in part provided by the Lawrencium computational cluster resource provided by the IT Division at the Lawrence Berkeley National Laboratory (supported by the Director, Office of Science, Office of Basic Energy Sciences, of the U.S. Department of Energy under Contract No. DE-AC02-05CH11231).
\end{acknowledgments}

\bibliography{refs.bib}

\begin{thebibliography}{63}%
\makeatletter
\providecommand \@ifxundefined [1]{%
 \@ifx{#1\undefined}
}%
\providecommand \@ifnum [1]{%
 \ifnum #1\expandafter \@firstoftwo
 \else \expandafter \@secondoftwo
 \fi
}%
\providecommand \@ifx [1]{%
 \ifx #1\expandafter \@firstoftwo
 \else \expandafter \@secondoftwo
 \fi
}%
\providecommand \natexlab [1]{#1}%
\providecommand \enquote  [1]{``#1''}%
\providecommand \bibnamefont  [1]{#1}%
\providecommand \bibfnamefont [1]{#1}%
\providecommand \citenamefont [1]{#1}%
\providecommand \href@noop [0]{\@secondoftwo}%
\providecommand \href [0]{\begingroup \@sanitize@url \@href}%
\providecommand \@href[1]{\@@startlink{#1}\@@href}%
\providecommand \@@href[1]{\endgroup#1\@@endlink}%
\providecommand \@sanitize@url [0]{\catcode `\\12\catcode `\$12\catcode `\&12\catcode `\#12\catcode `\^12\catcode `\_12\catcode `\%12\relax}%
\providecommand \@@startlink[1]{}%
\providecommand \@@endlink[0]{}%
\providecommand \url  [0]{\begingroup\@sanitize@url \@url }%
\providecommand \@url [1]{\endgroup\@href {#1}{\urlprefix }}%
\providecommand \urlprefix  [0]{URL }%
\providecommand \Eprint [0]{\href }%
\providecommand \doibase [0]{https://doi.org/}%
\providecommand \selectlanguage [0]{\@gobble}%
\providecommand \bibinfo  [0]{\@secondoftwo}%
\providecommand \bibfield  [0]{\@secondoftwo}%
\providecommand \translation [1]{[#1]}%
\providecommand \BibitemOpen [0]{}%
\providecommand \bibitemStop [0]{}%
\providecommand \bibitemNoStop [0]{.\EOS\space}%
\providecommand \EOS [0]{\spacefactor3000\relax}%
\providecommand \BibitemShut  [1]{\csname bibitem#1\endcsname}%
\let\auto@bib@innerbib\@empty
\bibitem [{\citenamefont {Kitaev}(2003)}]{kitaev2003fault}%
  \BibitemOpen
  \bibfield  {author} {\bibinfo {author} {\bibfnamefont {A.~Y.}\ \bibnamefont {Kitaev}},\ }\bibfield  {title} {\bibinfo {title} {Fault-tolerant quantum computation by anyons},\ }\href@noop {} {\bibfield  {journal} {\bibinfo  {journal} {Annals of physics}\ }\textbf {\bibinfo {volume} {303}},\ \bibinfo {pages} {2} (\bibinfo {year} {2003})}\BibitemShut {NoStop}%
\bibitem [{\citenamefont {Dennis}\ \emph {et~al.}(2002)\citenamefont {Dennis}, \citenamefont {Kitaev}, \citenamefont {Landahl},\ and\ \citenamefont {Preskill}}]{dennis2002topological}%
  \BibitemOpen
  \bibfield  {author} {\bibinfo {author} {\bibfnamefont {E.}~\bibnamefont {Dennis}}, \bibinfo {author} {\bibfnamefont {A.}~\bibnamefont {Kitaev}}, \bibinfo {author} {\bibfnamefont {A.}~\bibnamefont {Landahl}},\ and\ \bibinfo {author} {\bibfnamefont {J.}~\bibnamefont {Preskill}},\ }\bibfield  {title} {\bibinfo {title} {Topological quantum memory},\ }\href@noop {} {\bibfield  {journal} {\bibinfo  {journal} {Journal of Mathematical Physics}\ }\textbf {\bibinfo {volume} {43}},\ \bibinfo {pages} {4452} (\bibinfo {year} {2002})}\BibitemShut {NoStop}%
\bibitem [{\citenamefont {Semeghini}\ \emph {et~al.}(2021)\citenamefont {Semeghini}, \citenamefont {Levine}, \citenamefont {Keesling}, \citenamefont {Ebadi}, \citenamefont {Wang}, \citenamefont {Bluvstein}, \citenamefont {Verresen}, \citenamefont {Pichler}, \citenamefont {Kalinowski}, \citenamefont {Samajdar} \emph {et~al.}}]{semeghini2021probing}%
  \BibitemOpen
  \bibfield  {author} {\bibinfo {author} {\bibfnamefont {G.}~\bibnamefont {Semeghini}}, \bibinfo {author} {\bibfnamefont {H.}~\bibnamefont {Levine}}, \bibinfo {author} {\bibfnamefont {A.}~\bibnamefont {Keesling}}, \bibinfo {author} {\bibfnamefont {S.}~\bibnamefont {Ebadi}}, \bibinfo {author} {\bibfnamefont {T.~T.}\ \bibnamefont {Wang}}, \bibinfo {author} {\bibfnamefont {D.}~\bibnamefont {Bluvstein}}, \bibinfo {author} {\bibfnamefont {R.}~\bibnamefont {Verresen}}, \bibinfo {author} {\bibfnamefont {H.}~\bibnamefont {Pichler}}, \bibinfo {author} {\bibfnamefont {M.}~\bibnamefont {Kalinowski}}, \bibinfo {author} {\bibfnamefont {R.}~\bibnamefont {Samajdar}}, \emph {et~al.},\ }\bibfield  {title} {\bibinfo {title} {Probing topological spin liquids on a programmable quantum simulator},\ }\href@noop {} {\bibfield  {journal} {\bibinfo  {journal} {Science}\ }\textbf {\bibinfo {volume} {374}},\ \bibinfo {pages} {1242} (\bibinfo {year} {2021})}\BibitemShut {NoStop}%
\bibitem [{\citenamefont {Bluvstein}\ \emph {et~al.}(2022)\citenamefont {Bluvstein}, \citenamefont {Levine}, \citenamefont {Semeghini}, \citenamefont {Wang}, \citenamefont {Ebadi}, \citenamefont {Kalinowski}, \citenamefont {Keesling}, \citenamefont {Maskara}, \citenamefont {Pichler}, \citenamefont {Greiner} \emph {et~al.}}]{bluvstein2022quantum}%
  \BibitemOpen
  \bibfield  {author} {\bibinfo {author} {\bibfnamefont {D.}~\bibnamefont {Bluvstein}}, \bibinfo {author} {\bibfnamefont {H.}~\bibnamefont {Levine}}, \bibinfo {author} {\bibfnamefont {G.}~\bibnamefont {Semeghini}}, \bibinfo {author} {\bibfnamefont {T.~T.}\ \bibnamefont {Wang}}, \bibinfo {author} {\bibfnamefont {S.}~\bibnamefont {Ebadi}}, \bibinfo {author} {\bibfnamefont {M.}~\bibnamefont {Kalinowski}}, \bibinfo {author} {\bibfnamefont {A.}~\bibnamefont {Keesling}}, \bibinfo {author} {\bibfnamefont {N.}~\bibnamefont {Maskara}}, \bibinfo {author} {\bibfnamefont {H.}~\bibnamefont {Pichler}}, \bibinfo {author} {\bibfnamefont {M.}~\bibnamefont {Greiner}}, \emph {et~al.},\ }\bibfield  {title} {\bibinfo {title} {A quantum processor based on coherent transport of entangled atom arrays},\ }\href@noop {} {\bibfield  {journal} {\bibinfo  {journal} {Nature}\ }\textbf {\bibinfo {volume} {604}},\ \bibinfo {pages} {451} (\bibinfo {year} {2022})}\BibitemShut {NoStop}%
\bibitem [{\citenamefont {Satzinger}\ \emph {et~al.}(2021)\citenamefont {Satzinger}, \citenamefont {Liu}, \citenamefont {Smith}, \citenamefont {Knapp}, \citenamefont {Newman}, \citenamefont {Jones}, \citenamefont {Chen}, \citenamefont {Quintana}, \citenamefont {Mi}, \citenamefont {Dunsworth} \emph {et~al.}}]{satzinger2021realizing}%
  \BibitemOpen
  \bibfield  {author} {\bibinfo {author} {\bibfnamefont {K.}~\bibnamefont {Satzinger}}, \bibinfo {author} {\bibfnamefont {Y.-J.}\ \bibnamefont {Liu}}, \bibinfo {author} {\bibfnamefont {A.}~\bibnamefont {Smith}}, \bibinfo {author} {\bibfnamefont {C.}~\bibnamefont {Knapp}}, \bibinfo {author} {\bibfnamefont {M.}~\bibnamefont {Newman}}, \bibinfo {author} {\bibfnamefont {C.}~\bibnamefont {Jones}}, \bibinfo {author} {\bibfnamefont {Z.}~\bibnamefont {Chen}}, \bibinfo {author} {\bibfnamefont {C.}~\bibnamefont {Quintana}}, \bibinfo {author} {\bibfnamefont {X.}~\bibnamefont {Mi}}, \bibinfo {author} {\bibfnamefont {A.}~\bibnamefont {Dunsworth}}, \emph {et~al.},\ }\bibfield  {title} {\bibinfo {title} {Realizing topologically ordered states on a quantum processor},\ }\href@noop {} {\bibfield  {journal} {\bibinfo  {journal} {Science}\ }\textbf {\bibinfo {volume} {374}},\ \bibinfo {pages} {1237} (\bibinfo {year} {2021})}\BibitemShut {NoStop}%
\bibitem [{goo(2023)}]{google2023suppressing}%
  \BibitemOpen
  \bibfield  {title} {\bibinfo {title} {Suppressing quantum errors by scaling a surface code logical qubit},\ }\href@noop {} {\bibfield  {journal} {\bibinfo  {journal} {Nature}\ }\textbf {\bibinfo {volume} {614}},\ \bibinfo {pages} {676} (\bibinfo {year} {2023})}\BibitemShut {NoStop}%
\bibitem [{\citenamefont {Andersen}\ \emph {et~al.}(2023)\citenamefont {Andersen}, \citenamefont {Lensky}, \citenamefont {Kechedzhi}, \citenamefont {Drozdov}, \citenamefont {Bengtsson}, \citenamefont {Hong}, \citenamefont {Morvan}, \citenamefont {Mi}, \citenamefont {Opremcak}, \citenamefont {Kim} \emph {et~al.}}]{andersen2023observation}%
  \BibitemOpen
  \bibfield  {author} {\bibinfo {author} {\bibfnamefont {T.}~\bibnamefont {Andersen}}, \bibinfo {author} {\bibfnamefont {Y.}~\bibnamefont {Lensky}}, \bibinfo {author} {\bibfnamefont {K.}~\bibnamefont {Kechedzhi}}, \bibinfo {author} {\bibfnamefont {I.}~\bibnamefont {Drozdov}}, \bibinfo {author} {\bibfnamefont {A.}~\bibnamefont {Bengtsson}}, \bibinfo {author} {\bibfnamefont {S.}~\bibnamefont {Hong}}, \bibinfo {author} {\bibfnamefont {A.}~\bibnamefont {Morvan}}, \bibinfo {author} {\bibfnamefont {X.}~\bibnamefont {Mi}}, \bibinfo {author} {\bibfnamefont {A.}~\bibnamefont {Opremcak}}, \bibinfo {author} {\bibfnamefont {E.-A.}\ \bibnamefont {Kim}}, \emph {et~al.},\ }\bibfield  {title} {\bibinfo {title} {Observation of non-abelian exchange statistics on a superconducting processor},\ }\href@noop {} {\bibfield  {journal} {\bibinfo  {journal} {Bulletin of the American Physical Society}\ }\textbf {\bibinfo {volume} {68}} (\bibinfo {year} {2023})}\BibitemShut {NoStop}%
\bibitem [{\citenamefont {Bluvstein}\ \emph {et~al.}(2023)\citenamefont {Bluvstein}, \citenamefont {Evered}, \citenamefont {Geim}, \citenamefont {Li}, \citenamefont {Zhou}, \citenamefont {Manovitz}, \citenamefont {Ebadi}, \citenamefont {Cain}, \citenamefont {Kalinowski}, \citenamefont {Hangleiter} \emph {et~al.}}]{bluvstein2023logical}%
  \BibitemOpen
  \bibfield  {author} {\bibinfo {author} {\bibfnamefont {D.}~\bibnamefont {Bluvstein}}, \bibinfo {author} {\bibfnamefont {S.~J.}\ \bibnamefont {Evered}}, \bibinfo {author} {\bibfnamefont {A.~A.}\ \bibnamefont {Geim}}, \bibinfo {author} {\bibfnamefont {S.~H.}\ \bibnamefont {Li}}, \bibinfo {author} {\bibfnamefont {H.}~\bibnamefont {Zhou}}, \bibinfo {author} {\bibfnamefont {T.}~\bibnamefont {Manovitz}}, \bibinfo {author} {\bibfnamefont {S.}~\bibnamefont {Ebadi}}, \bibinfo {author} {\bibfnamefont {M.}~\bibnamefont {Cain}}, \bibinfo {author} {\bibfnamefont {M.}~\bibnamefont {Kalinowski}}, \bibinfo {author} {\bibfnamefont {D.}~\bibnamefont {Hangleiter}}, \emph {et~al.},\ }\bibfield  {title} {\bibinfo {title} {Logical quantum processor based on reconfigurable atom arrays},\ }\href@noop {} {\bibfield  {journal} {\bibinfo  {journal} {Nature}\ ,\ \bibinfo {pages} {1}} (\bibinfo {year} {2023})}\BibitemShut {NoStop}%
\bibitem [{\citenamefont {Nishimori}(1981)}]{nishimori1981internal}%
  \BibitemOpen
  \bibfield  {author} {\bibinfo {author} {\bibfnamefont {H.}~\bibnamefont {Nishimori}},\ }\bibfield  {title} {\bibinfo {title} {Internal energy, specific heat and correlation function of the bond-random ising model},\ }\href@noop {} {\bibfield  {journal} {\bibinfo  {journal} {Progress of Theoretical Physics}\ }\textbf {\bibinfo {volume} {66}},\ \bibinfo {pages} {1169} (\bibinfo {year} {1981})}\BibitemShut {NoStop}%
\bibitem [{Note1()}]{Note1}%
  \BibitemOpen
  \bibinfo {note} {The ML decoding we consider is sometimes referred to in literature as \protect \textit {degenerate} quantum ML (DQML) decoding~\cite {iyer2015hardness,deMarti_iOlius_2024}. Degeneracy is not found in classical codes and indicates that multiple error strings can give rise to the same error syndrome.}\BibitemShut {Stop}%
\bibitem [{\citenamefont {Terhal}(2015)}]{terhal2015quantum}%
  \BibitemOpen
  \bibfield  {author} {\bibinfo {author} {\bibfnamefont {B.~M.}\ \bibnamefont {Terhal}},\ }\bibfield  {title} {\bibinfo {title} {Quantum error correction for quantum memories},\ }\href@noop {} {\bibfield  {journal} {\bibinfo  {journal} {Reviews of Modern Physics}\ }\textbf {\bibinfo {volume} {87}},\ \bibinfo {pages} {307} (\bibinfo {year} {2015})}\BibitemShut {NoStop}%
\bibitem [{\citenamefont {Bravyi}\ \emph {et~al.}(2018)\citenamefont {Bravyi}, \citenamefont {Englbrecht}, \citenamefont {K{\"o}nig},\ and\ \citenamefont {Peard}}]{bravyi2018correcting}%
  \BibitemOpen
  \bibfield  {author} {\bibinfo {author} {\bibfnamefont {S.}~\bibnamefont {Bravyi}}, \bibinfo {author} {\bibfnamefont {M.}~\bibnamefont {Englbrecht}}, \bibinfo {author} {\bibfnamefont {R.}~\bibnamefont {K{\"o}nig}},\ and\ \bibinfo {author} {\bibfnamefont {N.}~\bibnamefont {Peard}},\ }\bibfield  {title} {\bibinfo {title} {Correcting coherent errors with surface codes},\ }\href@noop {} {\bibfield  {journal} {\bibinfo  {journal} {npj Quantum Information}\ }\textbf {\bibinfo {volume} {4}},\ \bibinfo {pages} {55} (\bibinfo {year} {2018})}\BibitemShut {NoStop}%
\bibitem [{\citenamefont {Ferris}\ and\ \citenamefont {Poulin}(2014)}]{ferris2014tensor}%
  \BibitemOpen
  \bibfield  {author} {\bibinfo {author} {\bibfnamefont {A.~J.}\ \bibnamefont {Ferris}}\ and\ \bibinfo {author} {\bibfnamefont {D.}~\bibnamefont {Poulin}},\ }\bibfield  {title} {\bibinfo {title} {Tensor networks and quantum error correction},\ }\href@noop {} {\bibfield  {journal} {\bibinfo  {journal} {Physical review letters}\ }\textbf {\bibinfo {volume} {113}},\ \bibinfo {pages} {030501} (\bibinfo {year} {2014})}\BibitemShut {NoStop}%
\bibitem [{\citenamefont {Bravyi}\ \emph {et~al.}(2014)\citenamefont {Bravyi}, \citenamefont {Suchara},\ and\ \citenamefont {Vargo}}]{bravyi2014efficient}%
  \BibitemOpen
  \bibfield  {author} {\bibinfo {author} {\bibfnamefont {S.}~\bibnamefont {Bravyi}}, \bibinfo {author} {\bibfnamefont {M.}~\bibnamefont {Suchara}},\ and\ \bibinfo {author} {\bibfnamefont {A.}~\bibnamefont {Vargo}},\ }\bibfield  {title} {\bibinfo {title} {Efficient algorithms for maximum likelihood decoding in the surface code},\ }\href@noop {} {\bibfield  {journal} {\bibinfo  {journal} {Physical Review A}\ }\textbf {\bibinfo {volume} {90}},\ \bibinfo {pages} {032326} (\bibinfo {year} {2014})}\BibitemShut {NoStop}%
\bibitem [{\citenamefont {Darmawan}\ and\ \citenamefont {Poulin}(2017)}]{Darmawan_2017}%
  \BibitemOpen
  \bibfield  {author} {\bibinfo {author} {\bibfnamefont {A.~S.}\ \bibnamefont {Darmawan}}\ and\ \bibinfo {author} {\bibfnamefont {D.}~\bibnamefont {Poulin}},\ }\bibfield  {title} {\bibinfo {title} {Tensor-network simulations of the surface code under realistic noise},\ }\bibfield  {journal} {\bibinfo  {journal} {Physical Review Letters}\ }\textbf {\bibinfo {volume} {119}},\ \href {https://doi.org/10.1103/physrevlett.119.040502} {10.1103/physrevlett.119.040502} (\bibinfo {year} {2017})\BibitemShut {NoStop}%
\bibitem [{\citenamefont {Darmawan}\ and\ \citenamefont {Poulin}(2018)}]{Darmawan_2018}%
  \BibitemOpen
  \bibfield  {author} {\bibinfo {author} {\bibfnamefont {A.~S.}\ \bibnamefont {Darmawan}}\ and\ \bibinfo {author} {\bibfnamefont {D.}~\bibnamefont {Poulin}},\ }\bibfield  {title} {\bibinfo {title} {Linear-time general decoding algorithm for the surface code},\ }\bibfield  {journal} {\bibinfo  {journal} {Physical Review E}\ }\textbf {\bibinfo {volume} {97}},\ \href {https://doi.org/10.1103/physreve.97.051302} {10.1103/physreve.97.051302} (\bibinfo {year} {2018})\BibitemShut {NoStop}%
\bibitem [{\citenamefont {Tuckett}\ \emph {et~al.}(2018)\citenamefont {Tuckett}, \citenamefont {Bartlett},\ and\ \citenamefont {Flammia}}]{tuckett2018ultrahigh}%
  \BibitemOpen
  \bibfield  {author} {\bibinfo {author} {\bibfnamefont {D.~K.}\ \bibnamefont {Tuckett}}, \bibinfo {author} {\bibfnamefont {S.~D.}\ \bibnamefont {Bartlett}},\ and\ \bibinfo {author} {\bibfnamefont {S.~T.}\ \bibnamefont {Flammia}},\ }\bibfield  {title} {\bibinfo {title} {Ultrahigh error threshold for surface codes with biased noise},\ }\href@noop {} {\bibfield  {journal} {\bibinfo  {journal} {Physical review letters}\ }\textbf {\bibinfo {volume} {120}},\ \bibinfo {pages} {050505} (\bibinfo {year} {2018})}\BibitemShut {NoStop}%
\bibitem [{\citenamefont {Tuckett}\ \emph {et~al.}(2019)\citenamefont {Tuckett}, \citenamefont {Darmawan}, \citenamefont {Chubb}, \citenamefont {Bravyi}, \citenamefont {Bartlett},\ and\ \citenamefont {Flammia}}]{tuckett2019tailoring}%
  \BibitemOpen
  \bibfield  {author} {\bibinfo {author} {\bibfnamefont {D.~K.}\ \bibnamefont {Tuckett}}, \bibinfo {author} {\bibfnamefont {A.~S.}\ \bibnamefont {Darmawan}}, \bibinfo {author} {\bibfnamefont {C.~T.}\ \bibnamefont {Chubb}}, \bibinfo {author} {\bibfnamefont {S.}~\bibnamefont {Bravyi}}, \bibinfo {author} {\bibfnamefont {S.~D.}\ \bibnamefont {Bartlett}},\ and\ \bibinfo {author} {\bibfnamefont {S.~T.}\ \bibnamefont {Flammia}},\ }\bibfield  {title} {\bibinfo {title} {Tailoring surface codes for highly biased noise},\ }\href@noop {} {\bibfield  {journal} {\bibinfo  {journal} {Physical Review X}\ }\textbf {\bibinfo {volume} {9}},\ \bibinfo {pages} {041031} (\bibinfo {year} {2019})}\BibitemShut {NoStop}%
\bibitem [{\citenamefont {Chubb}\ and\ \citenamefont {Flammia}(2021)}]{chubb2021statistical}%
  \BibitemOpen
  \bibfield  {author} {\bibinfo {author} {\bibfnamefont {C.~T.}\ \bibnamefont {Chubb}}\ and\ \bibinfo {author} {\bibfnamefont {S.~T.}\ \bibnamefont {Flammia}},\ }\bibfield  {title} {\bibinfo {title} {Statistical mechanical models for quantum codes with correlated noise},\ }\href@noop {} {\bibfield  {journal} {\bibinfo  {journal} {Annales de l’Institut Henri Poincar{\'e} D}\ }\textbf {\bibinfo {volume} {8}},\ \bibinfo {pages} {269} (\bibinfo {year} {2021})}\BibitemShut {NoStop}%
\bibitem [{\citenamefont {Chubb}(2021)}]{chubb2021general}%
  \BibitemOpen
  \bibfield  {author} {\bibinfo {author} {\bibfnamefont {C.~T.}\ \bibnamefont {Chubb}},\ }\bibfield  {title} {\bibinfo {title} {General tensor network decoding of 2d pauli codes},\ }\href@noop {} {\bibfield  {journal} {\bibinfo  {journal} {arXiv preprint arXiv:2101.04125}\ } (\bibinfo {year} {2021})}\BibitemShut {NoStop}%
\bibitem [{\citenamefont {Darmawan}(2024)}]{darmawan2024optimaladaptationsurfacecodedecoders}%
  \BibitemOpen
  \bibfield  {author} {\bibinfo {author} {\bibfnamefont {A.~S.}\ \bibnamefont {Darmawan}},\ }\href {https://arxiv.org/abs/2403.08706} {\bibinfo {title} {Optimal adaptation of surface-code decoders to local noise}} (\bibinfo {year} {2024}),\ \Eprint {https://arxiv.org/abs/2403.08706} {arXiv:2403.08706 [quant-ph]} \BibitemShut {NoStop}%
\bibitem [{\citenamefont {Behrends}\ \emph {et~al.}(2024)\citenamefont {Behrends}, \citenamefont {Venn},\ and\ \citenamefont {B{\'e}ri}}]{behrends2022surface}%
  \BibitemOpen
  \bibfield  {author} {\bibinfo {author} {\bibfnamefont {J.}~\bibnamefont {Behrends}}, \bibinfo {author} {\bibfnamefont {F.}~\bibnamefont {Venn}},\ and\ \bibinfo {author} {\bibfnamefont {B.}~\bibnamefont {B{\'e}ri}},\ }\bibfield  {title} {\bibinfo {title} {Surface codes, quantum circuits, and entanglement phases},\ }\href@noop {} {\bibfield  {journal} {\bibinfo  {journal} {Physical Review Research}\ }\textbf {\bibinfo {volume} {6}},\ \bibinfo {pages} {013137} (\bibinfo {year} {2024})}\BibitemShut {NoStop}%
\bibitem [{\citenamefont {Li}\ \emph {et~al.}(2018)\citenamefont {Li}, \citenamefont {Chen},\ and\ \citenamefont {Fisher}}]{li2018quantum}%
  \BibitemOpen
  \bibfield  {author} {\bibinfo {author} {\bibfnamefont {Y.}~\bibnamefont {Li}}, \bibinfo {author} {\bibfnamefont {X.}~\bibnamefont {Chen}},\ and\ \bibinfo {author} {\bibfnamefont {M.~P.}\ \bibnamefont {Fisher}},\ }\bibfield  {title} {\bibinfo {title} {Quantum zeno effect and the many-body entanglement transition},\ }\href@noop {} {\bibfield  {journal} {\bibinfo  {journal} {Physical Review B}\ }\textbf {\bibinfo {volume} {98}},\ \bibinfo {pages} {205136} (\bibinfo {year} {2018})}\BibitemShut {NoStop}%
\bibitem [{\citenamefont {Skinner}\ \emph {et~al.}(2019)\citenamefont {Skinner}, \citenamefont {Ruhman},\ and\ \citenamefont {Nahum}}]{skinner2019measurement}%
  \BibitemOpen
  \bibfield  {author} {\bibinfo {author} {\bibfnamefont {B.}~\bibnamefont {Skinner}}, \bibinfo {author} {\bibfnamefont {J.}~\bibnamefont {Ruhman}},\ and\ \bibinfo {author} {\bibfnamefont {A.}~\bibnamefont {Nahum}},\ }\bibfield  {title} {\bibinfo {title} {Measurement-induced phase transitions in the dynamics of entanglement},\ }\href@noop {} {\bibfield  {journal} {\bibinfo  {journal} {Physical Review X}\ }\textbf {\bibinfo {volume} {9}},\ \bibinfo {pages} {031009} (\bibinfo {year} {2019})}\BibitemShut {NoStop}%
\bibitem [{\citenamefont {Li}\ \emph {et~al.}(2019)\citenamefont {Li}, \citenamefont {Chen},\ and\ \citenamefont {Fisher}}]{li2019measurement}%
  \BibitemOpen
  \bibfield  {author} {\bibinfo {author} {\bibfnamefont {Y.}~\bibnamefont {Li}}, \bibinfo {author} {\bibfnamefont {X.}~\bibnamefont {Chen}},\ and\ \bibinfo {author} {\bibfnamefont {M.~P.}\ \bibnamefont {Fisher}},\ }\bibfield  {title} {\bibinfo {title} {Measurement-driven entanglement transition in hybrid quantum circuits},\ }\href@noop {} {\bibfield  {journal} {\bibinfo  {journal} {Physical Review B}\ }\textbf {\bibinfo {volume} {100}},\ \bibinfo {pages} {134306} (\bibinfo {year} {2019})}\BibitemShut {NoStop}%
\bibitem [{\citenamefont {Choi}\ \emph {et~al.}(2020)\citenamefont {Choi}, \citenamefont {Bao}, \citenamefont {Qi},\ and\ \citenamefont {Altman}}]{choi2020quantum}%
  \BibitemOpen
  \bibfield  {author} {\bibinfo {author} {\bibfnamefont {S.}~\bibnamefont {Choi}}, \bibinfo {author} {\bibfnamefont {Y.}~\bibnamefont {Bao}}, \bibinfo {author} {\bibfnamefont {X.-L.}\ \bibnamefont {Qi}},\ and\ \bibinfo {author} {\bibfnamefont {E.}~\bibnamefont {Altman}},\ }\bibfield  {title} {\bibinfo {title} {Quantum error correction in scrambling dynamics and measurement-induced phase transition},\ }\href@noop {} {\bibfield  {journal} {\bibinfo  {journal} {Physical Review Letters}\ }\textbf {\bibinfo {volume} {125}},\ \bibinfo {pages} {030505} (\bibinfo {year} {2020})}\BibitemShut {NoStop}%
\bibitem [{\citenamefont {Gullans}\ and\ \citenamefont {Huse}(2020)}]{gullans2020dynamical}%
  \BibitemOpen
  \bibfield  {author} {\bibinfo {author} {\bibfnamefont {M.~J.}\ \bibnamefont {Gullans}}\ and\ \bibinfo {author} {\bibfnamefont {D.~A.}\ \bibnamefont {Huse}},\ }\bibfield  {title} {\bibinfo {title} {Dynamical purification phase transition induced by quantum measurements},\ }\href@noop {} {\bibfield  {journal} {\bibinfo  {journal} {Physical Review X}\ }\textbf {\bibinfo {volume} {10}},\ \bibinfo {pages} {041020} (\bibinfo {year} {2020})}\BibitemShut {NoStop}%
\bibitem [{\citenamefont {Potter}\ and\ \citenamefont {Vasseur}(2022)}]{potter2022entanglement}%
  \BibitemOpen
  \bibfield  {author} {\bibinfo {author} {\bibfnamefont {A.~C.}\ \bibnamefont {Potter}}\ and\ \bibinfo {author} {\bibfnamefont {R.}~\bibnamefont {Vasseur}},\ }\bibfield  {title} {\bibinfo {title} {Entanglement dynamics in hybrid quantum circuits},\ }in\ \href@noop {} {\emph {\bibinfo {booktitle} {Entanglement in Spin Chains: From Theory to Quantum Technology Applications}}}\ (\bibinfo  {publisher} {Springer},\ \bibinfo {year} {2022})\ pp.\ \bibinfo {pages} {211--249}\BibitemShut {NoStop}%
\bibitem [{\citenamefont {Fisher}\ \emph {et~al.}(2023)\citenamefont {Fisher}, \citenamefont {Khemani}, \citenamefont {Nahum},\ and\ \citenamefont {Vijay}}]{fisher2023random}%
  \BibitemOpen
  \bibfield  {author} {\bibinfo {author} {\bibfnamefont {M.~P.}\ \bibnamefont {Fisher}}, \bibinfo {author} {\bibfnamefont {V.}~\bibnamefont {Khemani}}, \bibinfo {author} {\bibfnamefont {A.}~\bibnamefont {Nahum}},\ and\ \bibinfo {author} {\bibfnamefont {S.}~\bibnamefont {Vijay}},\ }\bibfield  {title} {\bibinfo {title} {Random quantum circuits},\ }\href@noop {} {\bibfield  {journal} {\bibinfo  {journal} {Annual Review of Condensed Matter Physics}\ }\textbf {\bibinfo {volume} {14}},\ \bibinfo {pages} {335} (\bibinfo {year} {2023})}\BibitemShut {NoStop}%
\bibitem [{\citenamefont {Sang}\ and\ \citenamefont {Hsieh}(2021)}]{sang2021measurement}%
  \BibitemOpen
  \bibfield  {author} {\bibinfo {author} {\bibfnamefont {S.}~\bibnamefont {Sang}}\ and\ \bibinfo {author} {\bibfnamefont {T.~H.}\ \bibnamefont {Hsieh}},\ }\bibfield  {title} {\bibinfo {title} {Measurement-protected quantum phases},\ }\href@noop {} {\bibfield  {journal} {\bibinfo  {journal} {Physical Review Research}\ }\textbf {\bibinfo {volume} {3}},\ \bibinfo {pages} {023200} (\bibinfo {year} {2021})}\BibitemShut {NoStop}%
\bibitem [{\citenamefont {Alberton}\ \emph {et~al.}(2021)\citenamefont {Alberton}, \citenamefont {Buchhold},\ and\ \citenamefont {Diehl}}]{alberton2021entanglement}%
  \BibitemOpen
  \bibfield  {author} {\bibinfo {author} {\bibfnamefont {O.}~\bibnamefont {Alberton}}, \bibinfo {author} {\bibfnamefont {M.}~\bibnamefont {Buchhold}},\ and\ \bibinfo {author} {\bibfnamefont {S.}~\bibnamefont {Diehl}},\ }\bibfield  {title} {\bibinfo {title} {Entanglement transition in a monitored free-fermion chain: From extended criticality to area law},\ }\href@noop {} {\bibfield  {journal} {\bibinfo  {journal} {Physical Review Letters}\ }\textbf {\bibinfo {volume} {126}},\ \bibinfo {pages} {170602} (\bibinfo {year} {2021})}\BibitemShut {NoStop}%
\bibitem [{\citenamefont {Zaletel}\ and\ \citenamefont {Pollmann}(2020)}]{zaletel2020isometric}%
  \BibitemOpen
  \bibfield  {author} {\bibinfo {author} {\bibfnamefont {M.~P.}\ \bibnamefont {Zaletel}}\ and\ \bibinfo {author} {\bibfnamefont {F.}~\bibnamefont {Pollmann}},\ }\bibfield  {title} {\bibinfo {title} {Isometric tensor network states in two dimensions},\ }\href@noop {} {\bibfield  {journal} {\bibinfo  {journal} {Physical review letters}\ }\textbf {\bibinfo {volume} {124}},\ \bibinfo {pages} {037201} (\bibinfo {year} {2020})}\BibitemShut {NoStop}%
\bibitem [{Note2()}]{Note2}%
  \BibitemOpen
  \bibinfo {note} {As discussed later, horizontal two-qubit (2Q) noise is sufficient to break the Gaussianity of the single-qubit X-type Pauli rotations. As (1) horizontal two-qubit noise is simpler to implement in our isometric tensor network sampling than vertical or nearest-neighbor 2Q errors and (2) we do not expect our results to qualitatively depend on the choice of non-Gaussianity, we restrict to horizontal 2Q errors.}\BibitemShut {Stop}%
\bibitem [{\citenamefont {Wang}\ \emph {et~al.}(2003)\citenamefont {Wang}, \citenamefont {Harrington},\ and\ \citenamefont {Preskill}}]{wang2003confinement}%
  \BibitemOpen
  \bibfield  {author} {\bibinfo {author} {\bibfnamefont {C.}~\bibnamefont {Wang}}, \bibinfo {author} {\bibfnamefont {J.}~\bibnamefont {Harrington}},\ and\ \bibinfo {author} {\bibfnamefont {J.}~\bibnamefont {Preskill}},\ }\bibfield  {title} {\bibinfo {title} {Confinement-higgs transition in a disordered gauge theory and the accuracy threshold for quantum memory},\ }\href@noop {} {\bibfield  {journal} {\bibinfo  {journal} {Annals of Physics}\ }\textbf {\bibinfo {volume} {303}},\ \bibinfo {pages} {31} (\bibinfo {year} {2003})}\BibitemShut {NoStop}%
\bibitem [{\citenamefont {Katzgraber}\ \emph {et~al.}(2009)\citenamefont {Katzgraber}, \citenamefont {Bomb{\'\i}n},\ and\ \citenamefont {Martin-Delgado}}]{katzgraber2009error}%
  \BibitemOpen
  \bibfield  {author} {\bibinfo {author} {\bibfnamefont {H.~G.}\ \bibnamefont {Katzgraber}}, \bibinfo {author} {\bibfnamefont {H.}~\bibnamefont {Bomb{\'\i}n}},\ and\ \bibinfo {author} {\bibfnamefont {M.~A.}\ \bibnamefont {Martin-Delgado}},\ }\bibfield  {title} {\bibinfo {title} {Error threshold for color codes and random three-body ising models},\ }\href@noop {} {\bibfield  {journal} {\bibinfo  {journal} {Physical review letters}\ }\textbf {\bibinfo {volume} {103}},\ \bibinfo {pages} {090501} (\bibinfo {year} {2009})}\BibitemShut {NoStop}%
\bibitem [{\citenamefont {Bombin}\ \emph {et~al.}(2012)\citenamefont {Bombin}, \citenamefont {Andrist}, \citenamefont {Ohzeki}, \citenamefont {Katzgraber},\ and\ \citenamefont {Martin-Delgado}}]{bombin2012strong}%
  \BibitemOpen
  \bibfield  {author} {\bibinfo {author} {\bibfnamefont {H.}~\bibnamefont {Bombin}}, \bibinfo {author} {\bibfnamefont {R.~S.}\ \bibnamefont {Andrist}}, \bibinfo {author} {\bibfnamefont {M.}~\bibnamefont {Ohzeki}}, \bibinfo {author} {\bibfnamefont {H.~G.}\ \bibnamefont {Katzgraber}},\ and\ \bibinfo {author} {\bibfnamefont {M.~A.}\ \bibnamefont {Martin-Delgado}},\ }\bibfield  {title} {\bibinfo {title} {Strong resilience of topological codes to depolarization},\ }\href@noop {} {\bibfield  {journal} {\bibinfo  {journal} {Physical Review X}\ }\textbf {\bibinfo {volume} {2}},\ \bibinfo {pages} {021004} (\bibinfo {year} {2012})}\BibitemShut {NoStop}%
\bibitem [{\citenamefont {Kubica}\ \emph {et~al.}(2018)\citenamefont {Kubica}, \citenamefont {Beverland}, \citenamefont {Brand{\~a}o}, \citenamefont {Preskill},\ and\ \citenamefont {Svore}}]{kubica2018three}%
  \BibitemOpen
  \bibfield  {author} {\bibinfo {author} {\bibfnamefont {A.}~\bibnamefont {Kubica}}, \bibinfo {author} {\bibfnamefont {M.~E.}\ \bibnamefont {Beverland}}, \bibinfo {author} {\bibfnamefont {F.}~\bibnamefont {Brand{\~a}o}}, \bibinfo {author} {\bibfnamefont {J.}~\bibnamefont {Preskill}},\ and\ \bibinfo {author} {\bibfnamefont {K.~M.}\ \bibnamefont {Svore}},\ }\bibfield  {title} {\bibinfo {title} {Three-dimensional color code thresholds via statistical-mechanical mapping},\ }\href@noop {} {\bibfield  {journal} {\bibinfo  {journal} {Physical review letters}\ }\textbf {\bibinfo {volume} {120}},\ \bibinfo {pages} {180501} (\bibinfo {year} {2018})}\BibitemShut {NoStop}%
\bibitem [{\citenamefont {Song}\ \emph {et~al.}(2022)\citenamefont {Song}, \citenamefont {Sch{\"o}nmeier-Kromer}, \citenamefont {Liu}, \citenamefont {Viyuela}, \citenamefont {Pollet},\ and\ \citenamefont {Martin-Delgado}}]{song2022optimal}%
  \BibitemOpen
  \bibfield  {author} {\bibinfo {author} {\bibfnamefont {H.}~\bibnamefont {Song}}, \bibinfo {author} {\bibfnamefont {J.}~\bibnamefont {Sch{\"o}nmeier-Kromer}}, \bibinfo {author} {\bibfnamefont {K.}~\bibnamefont {Liu}}, \bibinfo {author} {\bibfnamefont {O.}~\bibnamefont {Viyuela}}, \bibinfo {author} {\bibfnamefont {L.}~\bibnamefont {Pollet}},\ and\ \bibinfo {author} {\bibfnamefont {M.~A.}\ \bibnamefont {Martin-Delgado}},\ }\bibfield  {title} {\bibinfo {title} {Optimal thresholds for fracton codes and random spin models with subsystem symmetry},\ }\href@noop {} {\bibfield  {journal} {\bibinfo  {journal} {Physical Review Letters}\ }\textbf {\bibinfo {volume} {129}},\ \bibinfo {pages} {230502} (\bibinfo {year} {2022})}\BibitemShut {NoStop}%
\bibitem [{\citenamefont {Honecker}\ \emph {et~al.}(2001)\citenamefont {Honecker}, \citenamefont {Picco},\ and\ \citenamefont {Pujol}}]{honecker2001universality}%
  \BibitemOpen
  \bibfield  {author} {\bibinfo {author} {\bibfnamefont {A.}~\bibnamefont {Honecker}}, \bibinfo {author} {\bibfnamefont {M.}~\bibnamefont {Picco}},\ and\ \bibinfo {author} {\bibfnamefont {P.}~\bibnamefont {Pujol}},\ }\bibfield  {title} {\bibinfo {title} {Universality class of the nishimori point in the 2d$\pm$j random-bond ising model},\ }\href@noop {} {\bibfield  {journal} {\bibinfo  {journal} {Physical review letters}\ }\textbf {\bibinfo {volume} {87}},\ \bibinfo {pages} {047201} (\bibinfo {year} {2001})}\BibitemShut {NoStop}%
\bibitem [{\citenamefont {Venn}\ \emph {et~al.}(2023)\citenamefont {Venn}, \citenamefont {Behrends},\ and\ \citenamefont {B{\'e}ri}}]{venn2022coherent}%
  \BibitemOpen
  \bibfield  {author} {\bibinfo {author} {\bibfnamefont {F.}~\bibnamefont {Venn}}, \bibinfo {author} {\bibfnamefont {J.}~\bibnamefont {Behrends}},\ and\ \bibinfo {author} {\bibfnamefont {B.}~\bibnamefont {B{\'e}ri}},\ }\bibfield  {title} {\bibinfo {title} {Coherent-error threshold for surface codes from majorana delocalization},\ }\href@noop {} {\bibfield  {journal} {\bibinfo  {journal} {Physical Review Letters}\ }\textbf {\bibinfo {volume} {131}},\ \bibinfo {pages} {060603} (\bibinfo {year} {2023})}\BibitemShut {NoStop}%
\bibitem [{\citenamefont {Verstraete}\ and\ \citenamefont {Cirac}(2006)}]{verstraete2006matrix}%
  \BibitemOpen
  \bibfield  {author} {\bibinfo {author} {\bibfnamefont {F.}~\bibnamefont {Verstraete}}\ and\ \bibinfo {author} {\bibfnamefont {J.~I.}\ \bibnamefont {Cirac}},\ }\bibfield  {title} {\bibinfo {title} {Matrix product states represent ground states faithfully},\ }\href@noop {} {\bibfield  {journal} {\bibinfo  {journal} {Physical Review B—Condensed Matter and Materials Physics}\ }\textbf {\bibinfo {volume} {73}},\ \bibinfo {pages} {094423} (\bibinfo {year} {2006})}\BibitemShut {NoStop}%
\bibitem [{\citenamefont {Schuch}\ \emph {et~al.}(2008)\citenamefont {Schuch}, \citenamefont {Wolf}, \citenamefont {Verstraete},\ and\ \citenamefont {Cirac}}]{schuch2008entropy}%
  \BibitemOpen
  \bibfield  {author} {\bibinfo {author} {\bibfnamefont {N.}~\bibnamefont {Schuch}}, \bibinfo {author} {\bibfnamefont {M.~M.}\ \bibnamefont {Wolf}}, \bibinfo {author} {\bibfnamefont {F.}~\bibnamefont {Verstraete}},\ and\ \bibinfo {author} {\bibfnamefont {J.~I.}\ \bibnamefont {Cirac}},\ }\bibfield  {title} {\bibinfo {title} {Entropy scaling and simulability by matrix product states},\ }\href@noop {} {\bibfield  {journal} {\bibinfo  {journal} {Physical review letters}\ }\textbf {\bibinfo {volume} {100}},\ \bibinfo {pages} {030504} (\bibinfo {year} {2008})}\BibitemShut {NoStop}%
\bibitem [{SOM()}]{SOM}%
  \BibitemOpen
  \href@noop {} {}\bibinfo {note} {See supplementary online material for details.}\BibitemShut {Stop}%
\bibitem [{\citenamefont {Vidal}(2004)}]{vidal2004efficient}%
  \BibitemOpen
  \bibfield  {author} {\bibinfo {author} {\bibfnamefont {G.}~\bibnamefont {Vidal}},\ }\bibfield  {title} {\bibinfo {title} {Efficient simulation of one-dimensional quantum many-body systems},\ }\href@noop {} {\bibfield  {journal} {\bibinfo  {journal} {Physical review letters}\ }\textbf {\bibinfo {volume} {93}},\ \bibinfo {pages} {040502} (\bibinfo {year} {2004})}\BibitemShut {NoStop}%
\bibitem [{\citenamefont {Cao}\ \emph {et~al.}(2019)\citenamefont {Cao}, \citenamefont {Tilloy},\ and\ \citenamefont {De~Luca}}]{cao2019entanglement}%
  \BibitemOpen
  \bibfield  {author} {\bibinfo {author} {\bibfnamefont {X.}~\bibnamefont {Cao}}, \bibinfo {author} {\bibfnamefont {A.}~\bibnamefont {Tilloy}},\ and\ \bibinfo {author} {\bibfnamefont {A.}~\bibnamefont {De~Luca}},\ }\bibfield  {title} {\bibinfo {title} {Entanglement in a fermion chain under continuous monitoring},\ }\href@noop {} {\bibfield  {journal} {\bibinfo  {journal} {SciPost Physics}\ }\textbf {\bibinfo {volume} {7}},\ \bibinfo {pages} {024} (\bibinfo {year} {2019})}\BibitemShut {NoStop}%
\bibitem [{\citenamefont {Bao}\ \emph {et~al.}(2021)\citenamefont {Bao}, \citenamefont {Choi},\ and\ \citenamefont {Altman}}]{bao2021symmetry}%
  \BibitemOpen
  \bibfield  {author} {\bibinfo {author} {\bibfnamefont {Y.}~\bibnamefont {Bao}}, \bibinfo {author} {\bibfnamefont {S.}~\bibnamefont {Choi}},\ and\ \bibinfo {author} {\bibfnamefont {E.}~\bibnamefont {Altman}},\ }\bibfield  {title} {\bibinfo {title} {Symmetry enriched phases of quantum circuits},\ }\href@noop {} {\bibfield  {journal} {\bibinfo  {journal} {Annals of Physics}\ }\textbf {\bibinfo {volume} {435}},\ \bibinfo {pages} {168618} (\bibinfo {year} {2021})}\BibitemShut {NoStop}%
\bibitem [{\citenamefont {Jian}\ \emph {et~al.}(2022)\citenamefont {Jian}, \citenamefont {Bauer}, \citenamefont {Keselman},\ and\ \citenamefont {Ludwig}}]{jian2022criticality}%
  \BibitemOpen
  \bibfield  {author} {\bibinfo {author} {\bibfnamefont {C.-M.}\ \bibnamefont {Jian}}, \bibinfo {author} {\bibfnamefont {B.}~\bibnamefont {Bauer}}, \bibinfo {author} {\bibfnamefont {A.}~\bibnamefont {Keselman}},\ and\ \bibinfo {author} {\bibfnamefont {A.~W.}\ \bibnamefont {Ludwig}},\ }\bibfield  {title} {\bibinfo {title} {Criticality and entanglement in nonunitary quantum circuits and tensor networks of noninteracting fermions},\ }\href@noop {} {\bibfield  {journal} {\bibinfo  {journal} {Physical Review B}\ }\textbf {\bibinfo {volume} {106}},\ \bibinfo {pages} {134206} (\bibinfo {year} {2022})}\BibitemShut {NoStop}%
\bibitem [{\citenamefont {Fava}\ \emph {et~al.}(2023)\citenamefont {Fava}, \citenamefont {Piroli}, \citenamefont {Swann}, \citenamefont {Bernard},\ and\ \citenamefont {Nahum}}]{fava2023nonlinear}%
  \BibitemOpen
  \bibfield  {author} {\bibinfo {author} {\bibfnamefont {M.}~\bibnamefont {Fava}}, \bibinfo {author} {\bibfnamefont {L.}~\bibnamefont {Piroli}}, \bibinfo {author} {\bibfnamefont {T.}~\bibnamefont {Swann}}, \bibinfo {author} {\bibfnamefont {D.}~\bibnamefont {Bernard}},\ and\ \bibinfo {author} {\bibfnamefont {A.}~\bibnamefont {Nahum}},\ }\bibfield  {title} {\bibinfo {title} {Nonlinear sigma models for monitored dynamics of free fermions},\ }\href@noop {} {\bibfield  {journal} {\bibinfo  {journal} {Physical Review X}\ }\textbf {\bibinfo {volume} {13}},\ \bibinfo {pages} {041045} (\bibinfo {year} {2023})}\BibitemShut {NoStop}%
\bibitem [{\citenamefont {Jian}\ \emph {et~al.}(2023)\citenamefont {Jian}, \citenamefont {Shapourian}, \citenamefont {Bauer},\ and\ \citenamefont {Ludwig}}]{jian2023measurement}%
  \BibitemOpen
  \bibfield  {author} {\bibinfo {author} {\bibfnamefont {C.-M.}\ \bibnamefont {Jian}}, \bibinfo {author} {\bibfnamefont {H.}~\bibnamefont {Shapourian}}, \bibinfo {author} {\bibfnamefont {B.}~\bibnamefont {Bauer}},\ and\ \bibinfo {author} {\bibfnamefont {A.~W.}\ \bibnamefont {Ludwig}},\ }\bibfield  {title} {\bibinfo {title} {Measurement-induced entanglement transitions in quantum circuits of non-interacting fermions: Born-rule versus forced measurements},\ }\href@noop {} {\bibfield  {journal} {\bibinfo  {journal} {arXiv preprint arXiv:2302.09094}\ } (\bibinfo {year} {2023})}\BibitemShut {NoStop}%
\bibitem [{\citenamefont {Li}\ and\ \citenamefont {Fisher}(2023)}]{li2023decodable}%
  \BibitemOpen
  \bibfield  {author} {\bibinfo {author} {\bibfnamefont {Y.}~\bibnamefont {Li}}\ and\ \bibinfo {author} {\bibfnamefont {M.~P.}\ \bibnamefont {Fisher}},\ }\bibfield  {title} {\bibinfo {title} {Decodable hybrid dynamics of open quantum systems with z 2 symmetry},\ }\href@noop {} {\bibfield  {journal} {\bibinfo  {journal} {Physical Review B}\ }\textbf {\bibinfo {volume} {108}},\ \bibinfo {pages} {214302} (\bibinfo {year} {2023})}\BibitemShut {NoStop}%
\bibitem [{\citenamefont {Nielsen}\ and\ \citenamefont {Chuang}(2010)}]{nielsen2010quantum}%
  \BibitemOpen
  \bibfield  {author} {\bibinfo {author} {\bibfnamefont {M.~A.}\ \bibnamefont {Nielsen}}\ and\ \bibinfo {author} {\bibfnamefont {I.~L.}\ \bibnamefont {Chuang}},\ }\href@noop {} {\emph {\bibinfo {title} {Quantum computation and quantum information}}}\ (\bibinfo  {publisher} {Cambridge university press},\ \bibinfo {year} {2010})\BibitemShut {NoStop}%
\bibitem [{\citenamefont {Bao}\ \emph {et~al.}(2023)\citenamefont {Bao}, \citenamefont {Fan}, \citenamefont {Vishwanath},\ and\ \citenamefont {Altman}}]{bao2023mixed}%
  \BibitemOpen
  \bibfield  {author} {\bibinfo {author} {\bibfnamefont {Y.}~\bibnamefont {Bao}}, \bibinfo {author} {\bibfnamefont {R.}~\bibnamefont {Fan}}, \bibinfo {author} {\bibfnamefont {A.}~\bibnamefont {Vishwanath}},\ and\ \bibinfo {author} {\bibfnamefont {E.}~\bibnamefont {Altman}},\ }\bibfield  {title} {\bibinfo {title} {Mixed-state topological order and the errorfield double formulation of decoherence-induced transitions},\ }\href@noop {} {\bibfield  {journal} {\bibinfo  {journal} {arXiv preprint arXiv:2301.05687}\ } (\bibinfo {year} {2023})}\BibitemShut {NoStop}%
\bibitem [{\citenamefont {Fan}\ \emph {et~al.}(2023)\citenamefont {Fan}, \citenamefont {Bao}, \citenamefont {Altman},\ and\ \citenamefont {Vishwanath}}]{fan2023diagnostics}%
  \BibitemOpen
  \bibfield  {author} {\bibinfo {author} {\bibfnamefont {R.}~\bibnamefont {Fan}}, \bibinfo {author} {\bibfnamefont {Y.}~\bibnamefont {Bao}}, \bibinfo {author} {\bibfnamefont {E.}~\bibnamefont {Altman}},\ and\ \bibinfo {author} {\bibfnamefont {A.}~\bibnamefont {Vishwanath}},\ }\bibfield  {title} {\bibinfo {title} {Diagnostics of mixed-state topological order and breakdown of quantum memory},\ }\href@noop {} {\bibfield  {journal} {\bibinfo  {journal} {arXiv preprint arXiv:2301.05689}\ } (\bibinfo {year} {2023})}\BibitemShut {NoStop}%
\bibitem [{\citenamefont {Lee}\ \emph {et~al.}(2023)\citenamefont {Lee}, \citenamefont {Jian},\ and\ \citenamefont {Xu}}]{lee2023quantum}%
  \BibitemOpen
  \bibfield  {author} {\bibinfo {author} {\bibfnamefont {J.~Y.}\ \bibnamefont {Lee}}, \bibinfo {author} {\bibfnamefont {C.-M.}\ \bibnamefont {Jian}},\ and\ \bibinfo {author} {\bibfnamefont {C.}~\bibnamefont {Xu}},\ }\bibfield  {title} {\bibinfo {title} {Quantum criticality under decoherence or weak measurement},\ }\href@noop {} {\bibfield  {journal} {\bibinfo  {journal} {PRX quantum}\ }\textbf {\bibinfo {volume} {4}},\ \bibinfo {pages} {030317} (\bibinfo {year} {2023})}\BibitemShut {NoStop}%
\bibitem [{\citenamefont {Pollmann}\ \emph {et~al.}(2009)\citenamefont {Pollmann}, \citenamefont {Mukerjee}, \citenamefont {Turner},\ and\ \citenamefont {Moore}}]{pollmann2009theory}%
  \BibitemOpen
  \bibfield  {author} {\bibinfo {author} {\bibfnamefont {F.}~\bibnamefont {Pollmann}}, \bibinfo {author} {\bibfnamefont {S.}~\bibnamefont {Mukerjee}}, \bibinfo {author} {\bibfnamefont {A.~M.}\ \bibnamefont {Turner}},\ and\ \bibinfo {author} {\bibfnamefont {J.~E.}\ \bibnamefont {Moore}},\ }\bibfield  {title} {\bibinfo {title} {Theory of finite-entanglement scaling at one-dimensional quantum critical points},\ }\href@noop {} {\bibfield  {journal} {\bibinfo  {journal} {Physical review letters}\ }\textbf {\bibinfo {volume} {102}},\ \bibinfo {pages} {255701} (\bibinfo {year} {2009})}\BibitemShut {NoStop}%
\bibitem [{\citenamefont {Behrends}\ and\ \citenamefont {B{\'e}ri}(2024)}]{behrends2024statistical}%
  \BibitemOpen
  \bibfield  {author} {\bibinfo {author} {\bibfnamefont {J.}~\bibnamefont {Behrends}}\ and\ \bibinfo {author} {\bibfnamefont {B.}~\bibnamefont {B{\'e}ri}},\ }\bibfield  {title} {\bibinfo {title} {Statistical mechanical mapping and maximum-likelihood thresholds for the surface code under generic single-qubit coherent errors},\ }\href@noop {} {\bibfield  {journal} {\bibinfo  {journal} {arXiv preprint arXiv:2410.22436}\ } (\bibinfo {year} {2024})}\BibitemShut {NoStop}%
\bibitem [{\citenamefont {Cheng}\ \emph {et~al.}(2025)\citenamefont {Cheng}, \citenamefont {Huang}, \citenamefont {Khemani}, \citenamefont {Gullans},\ and\ \citenamefont {Ippoliti}}]{cheng2025emergent}%
  \BibitemOpen
  \bibfield  {author} {\bibinfo {author} {\bibfnamefont {Z.}~\bibnamefont {Cheng}}, \bibinfo {author} {\bibfnamefont {E.}~\bibnamefont {Huang}}, \bibinfo {author} {\bibfnamefont {V.}~\bibnamefont {Khemani}}, \bibinfo {author} {\bibfnamefont {M.~J.}\ \bibnamefont {Gullans}},\ and\ \bibinfo {author} {\bibfnamefont {M.}~\bibnamefont {Ippoliti}},\ }\bibfield  {title} {\bibinfo {title} {Emergent unitary designs for encoded qubits from coherent errors and syndrome measurements},\ }\href@noop {} {\bibfield  {journal} {\bibinfo  {journal} {PRX Quantum}\ }\textbf {\bibinfo {volume} {6}},\ \bibinfo {pages} {030333} (\bibinfo {year} {2025})}\BibitemShut {NoStop}%
\bibitem [{\citenamefont {Petravick}\ \emph {et~al.}(2007)\citenamefont {Petravick}, \citenamefont {Kramer}, \citenamefont {Olson}, \citenamefont {Livny}, \citenamefont {Roy}, \citenamefont {Avery}, \citenamefont {Blackburn}, \citenamefont {Wenaus}, \citenamefont {W{\"u}rthwein}, \citenamefont {Foster} \emph {et~al.}}]{osg1}%
  \BibitemOpen
  \bibfield  {author} {\bibinfo {author} {\bibfnamefont {D.}~\bibnamefont {Petravick}}, \bibinfo {author} {\bibfnamefont {B.}~\bibnamefont {Kramer}}, \bibinfo {author} {\bibfnamefont {D.}~\bibnamefont {Olson}}, \bibinfo {author} {\bibfnamefont {M.}~\bibnamefont {Livny}}, \bibinfo {author} {\bibfnamefont {A.}~\bibnamefont {Roy}}, \bibinfo {author} {\bibfnamefont {P.}~\bibnamefont {Avery}}, \bibinfo {author} {\bibfnamefont {K.}~\bibnamefont {Blackburn}}, \bibinfo {author} {\bibfnamefont {T.}~\bibnamefont {Wenaus}}, \bibinfo {author} {\bibfnamefont {F.}~\bibnamefont {W{\"u}rthwein}}, \bibinfo {author} {\bibfnamefont {I.}~\bibnamefont {Foster}}, \emph {et~al.},\ }\bibfield  {title} {\bibinfo {title} {The open science grid},\ }in\ \href@noop {} {\emph {\bibinfo {booktitle} {Journal of Physics: Conference Series}}},\ Vol.~\bibinfo {volume} {78}\ (\bibinfo {organization} {IOP Publishing},\ \bibinfo {year} {2007})\ p.\ \bibinfo {pages} {012057}\BibitemShut {NoStop}%
\bibitem [{\citenamefont {Sfiligoi}\ \emph {et~al.}(2009)\citenamefont {Sfiligoi}, \citenamefont {Bradley}, \citenamefont {Holzman}, \citenamefont {Mhashilkar}, \citenamefont {Padhi},\ and\ \citenamefont {Wurthwein}}]{osg2}%
  \BibitemOpen
  \bibfield  {author} {\bibinfo {author} {\bibfnamefont {I.}~\bibnamefont {Sfiligoi}}, \bibinfo {author} {\bibfnamefont {D.~C.}\ \bibnamefont {Bradley}}, \bibinfo {author} {\bibfnamefont {B.}~\bibnamefont {Holzman}}, \bibinfo {author} {\bibfnamefont {P.}~\bibnamefont {Mhashilkar}}, \bibinfo {author} {\bibfnamefont {S.}~\bibnamefont {Padhi}},\ and\ \bibinfo {author} {\bibfnamefont {F.}~\bibnamefont {Wurthwein}},\ }\bibfield  {title} {\bibinfo {title} {The pilot way to grid resources using glideinwms},\ }in\ \href@noop {} {\emph {\bibinfo {booktitle} {2009 WRI World congress on computer science and information engineering}}},\ Vol.~\bibinfo {volume} {2}\ (\bibinfo {organization} {IEEE},\ \bibinfo {year} {2009})\ pp.\ \bibinfo {pages} {428--432}\BibitemShut {NoStop}%
\bibitem [{\citenamefont {OSG}(2006)}]{osg3}%
  \BibitemOpen
  \bibfield  {author} {\bibinfo {author} {\bibnamefont {OSG}},\ }\bibfield  {title} {\bibinfo {title} {Ospool},\ }\bibfield  {journal} {\bibinfo  {journal} {OSG}\ }\href {https://doi.org/10.21231/906P-4D78} {10.21231/906P-4D78} (\bibinfo {year} {2006})\BibitemShut {NoStop}%
\bibitem [{\citenamefont {OSG}(2015)}]{osg4}%
  \BibitemOpen
  \bibfield  {author} {\bibinfo {author} {\bibnamefont {OSG}},\ }\bibfield  {title} {\bibinfo {title} {Open science data federation},\ }\bibfield  {journal} {\bibinfo  {journal} {OSG}\ }\href {https://doi.org/10.21231/0KVZ-VE57} {10.21231/0KVZ-VE57} (\bibinfo {year} {2015})\BibitemShut {NoStop}%
\bibitem [{\citenamefont {Iyer}\ and\ \citenamefont {Poulin}(2015)}]{iyer2015hardness}%
  \BibitemOpen
  \bibfield  {author} {\bibinfo {author} {\bibfnamefont {P.}~\bibnamefont {Iyer}}\ and\ \bibinfo {author} {\bibfnamefont {D.}~\bibnamefont {Poulin}},\ }\bibfield  {title} {\bibinfo {title} {Hardness of decoding quantum stabilizer codes},\ }\href@noop {} {\bibfield  {journal} {\bibinfo  {journal} {IEEE Transactions on Information Theory}\ }\textbf {\bibinfo {volume} {61}},\ \bibinfo {pages} {5209} (\bibinfo {year} {2015})}\BibitemShut {NoStop}%
\bibitem [{\citenamefont {deMarti iOlius}\ \emph {et~al.}(2024)\citenamefont {deMarti iOlius}, \citenamefont {Fuentes}, \citenamefont {Orús}, \citenamefont {Crespo},\ and\ \citenamefont {Etxezarreta~Martinez}}]{deMarti_iOlius_2024}%
  \BibitemOpen
  \bibfield  {author} {\bibinfo {author} {\bibfnamefont {A.}~\bibnamefont {deMarti iOlius}}, \bibinfo {author} {\bibfnamefont {P.}~\bibnamefont {Fuentes}}, \bibinfo {author} {\bibfnamefont {R.}~\bibnamefont {Orús}}, \bibinfo {author} {\bibfnamefont {P.~M.}\ \bibnamefont {Crespo}},\ and\ \bibinfo {author} {\bibfnamefont {J.}~\bibnamefont {Etxezarreta~Martinez}},\ }\bibfield  {title} {\bibinfo {title} {Decoding algorithms for surface codes},\ }\href {https://doi.org/10.22331/q-2024-10-10-1498} {\bibfield  {journal} {\bibinfo  {journal} {Quantum}\ }\textbf {\bibinfo {volume} {8}},\ \bibinfo {pages} {1498} (\bibinfo {year} {2024})}\BibitemShut {NoStop}%
\end{thebibliography}%


\begin{thebibliography}{14}%
\makeatletter
\providecommand \@ifxundefined [1]{%
 \@ifx{#1\undefined}
}%
\providecommand \@ifnum [1]{%
 \ifnum #1\expandafter \@firstoftwo
 \else \expandafter \@secondoftwo
 \fi
}%
\providecommand \@ifx [1]{%
 \ifx #1\expandafter \@firstoftwo
 \else \expandafter \@secondoftwo
 \fi
}%
\providecommand \natexlab [1]{#1}%
\providecommand \enquote  [1]{``#1''}%
\providecommand \bibnamefont  [1]{#1}%
\providecommand \bibfnamefont [1]{#1}%
\providecommand \citenamefont [1]{#1}%
\providecommand \href@noop [0]{\@secondoftwo}%
\providecommand \href [0]{\begingroup \@sanitize@url \@href}%
\providecommand \@href[1]{\@@startlink{#1}\@@href}%
\providecommand \@@href[1]{\endgroup#1\@@endlink}%
\providecommand \@sanitize@url [0]{\catcode `\\12\catcode `\$12\catcode `\&12\catcode `\#12\catcode `\^12\catcode `\_12\catcode `\%12\relax}%
\providecommand \@@startlink[1]{}%
\providecommand \@@endlink[0]{}%
\providecommand \url  [0]{\begingroup\@sanitize@url \@url }%
\providecommand \@url [1]{\endgroup\@href {#1}{\urlprefix }}%
\providecommand \urlprefix  [0]{URL }%
\providecommand \Eprint [0]{\href }%
\providecommand \doibase [0]{http://dx.doi.org/}%
\providecommand \selectlanguage [0]{\@gobble}%
\providecommand \bibinfo  [0]{\@secondoftwo}%
\providecommand \bibfield  [0]{\@secondoftwo}%
\providecommand \translation [1]{[#1]}%
\providecommand \BibitemOpen [0]{}%
\providecommand \bibitemStop [0]{}%
\providecommand \bibitemNoStop [0]{.\EOS\space}%
\providecommand \EOS [0]{\spacefactor3000\relax}%
\providecommand \BibitemShut  [1]{\csname bibitem#1\endcsname}%
\let\auto@bib@innerbib\@empty
\bibitem [{\citenamefont {Venn}\ \emph {et~al.}(2023)\citenamefont {Venn}, \citenamefont {Behrends},\ and\ \citenamefont {B{\'e}ri}}]{venn2022coherent}%
  \BibitemOpen
  \bibfield  {author} {\bibinfo {author} {\bibfnamefont {F.}~\bibnamefont {Venn}}, \bibinfo {author} {\bibfnamefont {J.}~\bibnamefont {Behrends}}, \ and\ \bibinfo {author} {\bibfnamefont {B.}~\bibnamefont {B{\'e}ri}},\ }\href@noop {} {\bibfield  {journal} {\bibinfo  {journal} {Physical Review Letters}\ }\textbf {\bibinfo {volume} {131}},\ \bibinfo {pages} {060603} (\bibinfo {year} {2023})}\BibitemShut {NoStop}%
\bibitem [{\citenamefont {Behrends}\ \emph {et~al.}(2024)\citenamefont {Behrends}, \citenamefont {Venn},\ and\ \citenamefont {B{\'e}ri}}]{behrends2022surface}%
  \BibitemOpen
  \bibfield  {author} {\bibinfo {author} {\bibfnamefont {J.}~\bibnamefont {Behrends}}, \bibinfo {author} {\bibfnamefont {F.}~\bibnamefont {Venn}}, \ and\ \bibinfo {author} {\bibfnamefont {B.}~\bibnamefont {B{\'e}ri}},\ }\href@noop {} {\bibfield  {journal} {\bibinfo  {journal} {Physical Review Research}\ }\textbf {\bibinfo {volume} {6}},\ \bibinfo {pages} {013137} (\bibinfo {year} {2024})}\BibitemShut {NoStop}%
\bibitem [{Note1()}]{Note1}%
  \BibitemOpen
  \bibinfo {note} {Note that the partition function involves a summation over the Boltzmann weights that are in general complex when considering coherent errors.}\BibitemShut {Stop}%
\bibitem [{\citenamefont {Zaletel}\ and\ \citenamefont {Pollmann}(2020)}]{zaletel2020isometric}%
  \BibitemOpen
  \bibfield  {author} {\bibinfo {author} {\bibfnamefont {M.~P.}\ \bibnamefont {Zaletel}}\ and\ \bibinfo {author} {\bibfnamefont {F.}~\bibnamefont {Pollmann}},\ }\href@noop {} {\bibfield  {journal} {\bibinfo  {journal} {Physical review letters}\ }\textbf {\bibinfo {volume} {124}},\ \bibinfo {pages} {037201} (\bibinfo {year} {2020})}\BibitemShut {NoStop}%
\bibitem [{\citenamefont {Lin}\ \emph {et~al.}(2022)\citenamefont {Lin}, \citenamefont {Zaletel},\ and\ \citenamefont {Pollmann}}]{Lin_2022}%
  \BibitemOpen
  \bibfield  {author} {\bibinfo {author} {\bibfnamefont {S.-H.}\ \bibnamefont {Lin}}, \bibinfo {author} {\bibfnamefont {M.~P.}\ \bibnamefont {Zaletel}}, \ and\ \bibinfo {author} {\bibfnamefont {F.}~\bibnamefont {Pollmann}},\ }\href {\doibase 10.1103/physrevb.106.245102} {\bibfield  {journal} {\bibinfo  {journal} {Physical Review B}\ }\textbf {\bibinfo {volume} {106}} (\bibinfo {year} {2022}),\ 10.1103/physrevb.106.245102}\BibitemShut {NoStop}%
\bibitem [{\citenamefont {Dai}\ \emph {et~al.}(2024)\citenamefont {Dai}, \citenamefont {Wu}, \citenamefont {Wang},\ and\ \citenamefont {Zaletel}}]{dai2024fermionicisometrictensornetwork}%
  \BibitemOpen
  \bibfield  {author} {\bibinfo {author} {\bibfnamefont {Z.}~\bibnamefont {Dai}}, \bibinfo {author} {\bibfnamefont {Y.}~\bibnamefont {Wu}}, \bibinfo {author} {\bibfnamefont {T.}~\bibnamefont {Wang}}, \ and\ \bibinfo {author} {\bibfnamefont {M.~P.}\ \bibnamefont {Zaletel}},\ }\href {https://arxiv.org/abs/2211.00043} {\enquote {\bibinfo {title} {Fermionic isometric tensor network states in two dimensions},}\ } (\bibinfo {year} {2024}),\ \Eprint {http://arxiv.org/abs/2211.00043} {arXiv:2211.00043 [cond-mat.str-el]} \BibitemShut {NoStop}%
\bibitem [{\citenamefont {Kadow}\ \emph {et~al.}(2023)\citenamefont {Kadow}, \citenamefont {Pollmann},\ and\ \citenamefont {Knap}}]{Kadow_2023}%
  \BibitemOpen
  \bibfield  {author} {\bibinfo {author} {\bibfnamefont {W.}~\bibnamefont {Kadow}}, \bibinfo {author} {\bibfnamefont {F.}~\bibnamefont {Pollmann}}, \ and\ \bibinfo {author} {\bibfnamefont {M.}~\bibnamefont {Knap}},\ }\href {\doibase 10.1103/physrevb.107.205106} {\bibfield  {journal} {\bibinfo  {journal} {Physical Review B}\ }\textbf {\bibinfo {volume} {107}} (\bibinfo {year} {2023}),\ 10.1103/physrevb.107.205106}\BibitemShut {NoStop}%
\bibitem [{\citenamefont {Soejima}\ \emph {et~al.}(2020)\citenamefont {Soejima}, \citenamefont {Siva}, \citenamefont {Bultinck}, \citenamefont {Chatterjee}, \citenamefont {Pollmann},\ and\ \citenamefont {Zaletel}}]{Soejima_2020}%
  \BibitemOpen
  \bibfield  {author} {\bibinfo {author} {\bibfnamefont {T.}~\bibnamefont {Soejima}}, \bibinfo {author} {\bibfnamefont {K.}~\bibnamefont {Siva}}, \bibinfo {author} {\bibfnamefont {N.}~\bibnamefont {Bultinck}}, \bibinfo {author} {\bibfnamefont {S.}~\bibnamefont {Chatterjee}}, \bibinfo {author} {\bibfnamefont {F.}~\bibnamefont {Pollmann}}, \ and\ \bibinfo {author} {\bibfnamefont {M.~P.}\ \bibnamefont {Zaletel}},\ }\href {\doibase 10.1103/physrevb.101.085117} {\bibfield  {journal} {\bibinfo  {journal} {Physical Review B}\ }\textbf {\bibinfo {volume} {101}} (\bibinfo {year} {2020}),\ 10.1103/physrevb.101.085117}\BibitemShut {NoStop}%
\bibitem [{\citenamefont {Malz}\ and\ \citenamefont {Trivedi}(2024)}]{malz2024computationalcomplexityisometrictensor}%
  \BibitemOpen
  \bibfield  {author} {\bibinfo {author} {\bibfnamefont {D.}~\bibnamefont {Malz}}\ and\ \bibinfo {author} {\bibfnamefont {R.}~\bibnamefont {Trivedi}},\ }\href {https://arxiv.org/abs/2402.07975} {\enquote {\bibinfo {title} {Computational complexity of isometric tensor network states},}\ } (\bibinfo {year} {2024}),\ \Eprint {http://arxiv.org/abs/2402.07975} {arXiv:2402.07975 [quant-ph]} \BibitemShut {NoStop}%
\bibitem [{\citenamefont {Liu}\ \emph {et~al.}(2024)\citenamefont {Liu}, \citenamefont {Shtengel},\ and\ \citenamefont {Pollmann}}]{liu2024simulating2dtopologicalquantum}%
  \BibitemOpen
  \bibfield  {author} {\bibinfo {author} {\bibfnamefont {Y.-J.}\ \bibnamefont {Liu}}, \bibinfo {author} {\bibfnamefont {K.}~\bibnamefont {Shtengel}}, \ and\ \bibinfo {author} {\bibfnamefont {F.}~\bibnamefont {Pollmann}},\ }\href {https://arxiv.org/abs/2312.05079} {\enquote {\bibinfo {title} {Simulating 2d topological quantum phase transitions on a digital quantum computer},}\ } (\bibinfo {year} {2024}),\ \Eprint {http://arxiv.org/abs/2312.05079} {arXiv:2312.05079 [quant-ph]} \BibitemShut {NoStop}%
\bibitem [{\citenamefont {Schön}\ \emph {et~al.}(2005)\citenamefont {Schön}, \citenamefont {Solano}, \citenamefont {Verstraete}, \citenamefont {Cirac},\ and\ \citenamefont {Wolf}}]{Sch_n_2005}%
  \BibitemOpen
  \bibfield  {author} {\bibinfo {author} {\bibfnamefont {C.}~\bibnamefont {Schön}}, \bibinfo {author} {\bibfnamefont {E.}~\bibnamefont {Solano}}, \bibinfo {author} {\bibfnamefont {F.}~\bibnamefont {Verstraete}}, \bibinfo {author} {\bibfnamefont {J.~I.}\ \bibnamefont {Cirac}}, \ and\ \bibinfo {author} {\bibfnamefont {M.~M.}\ \bibnamefont {Wolf}},\ }\href {\doibase 10.1103/physrevlett.95.110503} {\bibfield  {journal} {\bibinfo  {journal} {Physical Review Letters}\ }\textbf {\bibinfo {volume} {95}} (\bibinfo {year} {2005}),\ 10.1103/physrevlett.95.110503}\BibitemShut {NoStop}%
\bibitem [{\citenamefont {Bañuls}\ \emph {et~al.}(2008)\citenamefont {Bañuls}, \citenamefont {Pérez-García}, \citenamefont {Wolf}, \citenamefont {Verstraete},\ and\ \citenamefont {Cirac}}]{Ba_uls_2008}%
  \BibitemOpen
  \bibfield  {author} {\bibinfo {author} {\bibfnamefont {M.~C.}\ \bibnamefont {Bañuls}}, \bibinfo {author} {\bibfnamefont {D.}~\bibnamefont {Pérez-García}}, \bibinfo {author} {\bibfnamefont {M.~M.}\ \bibnamefont {Wolf}}, \bibinfo {author} {\bibfnamefont {F.}~\bibnamefont {Verstraete}}, \ and\ \bibinfo {author} {\bibfnamefont {J.~I.}\ \bibnamefont {Cirac}},\ }\href {\doibase 10.1103/physreva.77.052306} {\bibfield  {journal} {\bibinfo  {journal} {Physical Review A}\ }\textbf {\bibinfo {volume} {77}} (\bibinfo {year} {2008}),\ 10.1103/physreva.77.052306}\BibitemShut {NoStop}%
\bibitem [{\citenamefont {Wei}\ \emph {et~al.}(2022)\citenamefont {Wei}, \citenamefont {Malz},\ and\ \citenamefont {Cirac}}]{Wei_2022}%
  \BibitemOpen
  \bibfield  {author} {\bibinfo {author} {\bibfnamefont {Z.-Y.}\ \bibnamefont {Wei}}, \bibinfo {author} {\bibfnamefont {D.}~\bibnamefont {Malz}}, \ and\ \bibinfo {author} {\bibfnamefont {J.~I.}\ \bibnamefont {Cirac}},\ }\href {\doibase 10.1103/physrevlett.128.010607} {\bibfield  {journal} {\bibinfo  {journal} {Physical Review Letters}\ }\textbf {\bibinfo {volume} {128}} (\bibinfo {year} {2022}),\ 10.1103/physrevlett.128.010607}\BibitemShut {NoStop}%
\bibitem [{\citenamefont {Anand}\ \emph {et~al.}(2023)\citenamefont {Anand}, \citenamefont {Hauschild}, \citenamefont {Zhang}, \citenamefont {Potter},\ and\ \citenamefont {Zaletel}}]{Anand_2023}%
  \BibitemOpen
  \bibfield  {author} {\bibinfo {author} {\bibfnamefont {S.}~\bibnamefont {Anand}}, \bibinfo {author} {\bibfnamefont {J.}~\bibnamefont {Hauschild}}, \bibinfo {author} {\bibfnamefont {Y.}~\bibnamefont {Zhang}}, \bibinfo {author} {\bibfnamefont {A.~C.}\ \bibnamefont {Potter}}, \ and\ \bibinfo {author} {\bibfnamefont {M.~P.}\ \bibnamefont {Zaletel}},\ }\href {\doibase 10.1103/prxquantum.4.030334} {\bibfield  {journal} {\bibinfo  {journal} {PRX Quantum}\ }\textbf {\bibinfo {volume} {4}} (\bibinfo {year} {2023}),\ 10.1103/prxquantum.4.030334}\BibitemShut {NoStop}%
\end{thebibliography}%

\end{document}


\title{Supplementary Material for ``Phases of decodability in the surface code with unitary errors''}

\author{Yimu Bao}
\affiliation{Kavli Institute for Theoretical Physics, University of California, Santa Barbara, CA 93106, USA}

\author{Sajant Anand}
\affiliation{Department of Physics, University of California, Berkeley, CA 94720, USA}

\maketitle

\tableofcontents

\section{Statistical mechanics models for maximum likelihood decoding}
In this section, we derive the statistical mechanics (stat-mech) model associated with the maximum likelihood (ML) decoder for the surface code under unitary errors studied in this work.
We explicitly write down the tensor network representation for the partition function of stat-mech models and express the partition function in terms of the transfer matrices in one spatial direction.
Section~\ref{sec:coh-X} reviews the mapping for the surface code under Pauli-X rotation.
Section~\ref{sec:coh-X_coh-XX} develops the stat-mech model for the surface code under both single- and two-qubit Pauli-X rotations.
Section~\ref{sec:coh-XYZ_coh-XX} concerns the general single-qubit unitary rotation together with two-qubit Pauli-X rotation and details the stat-mech mapping.
Technically, the stat-mech model in Sec.~\ref{sec:coh-XYZ_coh-XX} includes the ones derived in Sec.~\ref{sec:coh-X} and~\ref{sec:coh-X_coh-XX} as special cases.
However, it involves additional degrees of freedom that need not be simulated in the decoding problem that only involve Pauli-X errors.

\subsection{X rotation}\label{sec:coh-X}
Here, we review the RBIM derived for the ML decoder in the surface code under Pauli-X rotations~\cite{venn2022coherent,behrends2022surface}.
The corrupted state $\ket{\Psi}$ takes the form
\begin{align}
    \ket{\Psi} = \prod_\ell e^{i \theta X_\ell} \ket{\Psi_0} = (\cos\theta)^N \sum_{\calC}(\ri\tan\theta)^{|\calC|}X^{\calC}\ket{\Psi_0},
\end{align}
where the encoded state $\ket{\Psi_0}$ is an eigenstate of logical-Z operator, and $N$ denotes total number of qubits.

To recover the encoded state, we first performs the syndrome measurements.
Based on the observed syndrome $s$, one can divide the error strings $\calC$ that are compatible with the syndrome (i.e. $\partial C = s$) into two homological classes.
The strings in two different classes differ by a logical $X$ operator along a non-contractible path.
The probability of error strings in each class is given by
\begin{align}
    P_{\sfa,s} = \left|\bra{\Psi_0}X^{\calR_\sfa} \hat{\Pi}_s \ket{\Psi}\right|^2.
\end{align}
where $\hat{\Pi}_s$ is the projection onto the space with syndrome $s$, and $\calR_\sfa$ is a reference error string in class $\sfa$ that is chosen to remove the syndrome.
In the ML decoder, we choose a recovery string operator from the class $\sfa$ with the highest probability to restore the encoded state.
We note that the choice of the recovery string $\calR_\sfa$ in class $\sfa$ is arbitrary and does not affect the resulting state after correction.

The probability $P_{\sfa,s}$ maps to the partition function of the random bond Ising model\footnote{Note that the partition function involves a summation over the Boltzmann weights that are in general complex when considering coherent errors.}, i.e. $P_{\sfa,s} = |\calZ_{\sfa,s}|^2$, with 
\begin{align}
    \calZ_{\sfa,s} = \sum_{\calC, [\calC] = \sfa} (\cos\theta)^{N-|\calC|} (\ri \sin\theta)^{|\calC|},
\end{align} 
where $\calC$ is an error configuration on the dual lattice.
Each string $\calC_\sfa$ in class $\sfa$ can be written as $\calC_\sfa = calR_\sfa + \partial v$, a combination of the reference string, and a contractible loop $\partial v$ given by the boundary of a collection of vertices $v$.
Here, we introduce random bond variables $\eta_{ij}$ taking $-1$ values along the reference string $\calR_\sfa$. an Ising spin $\sigma_i$ on each vertex.
We use the domain wall of $\sigma_i$ to represent the contractible loop $\partial v$.
In this way, we have
\begin{align}
    |\calC_\sfa|_{ij} = \frac{1 - \eta_{ij}\sigma_i\sigma_j}{2},.\label{eq:error_string_spin}
\end{align}

The partition function in terms of the Ising variables takes the form $\calZ_{\sfa,s} = \sum_{\{\sigma_i\}} e^{-H(\{\sigma_i\})}$ with the Hamiltonian given by
\begin{align}
    H = \sum_{\langle i,j\rangle}-\left(J_\theta - \frac{\ri \pi}{4}\right) \eta_{ij} \sigma_i \sigma_j,
\end{align}
where $J_\theta = (1/2)\log(1/\tan\theta)$, and we have neglected an unimportant constant term in the Hamiltonian.
The RBIM has open boundary conditions in the vertical direction, while it has symmetry-breaking boundary conditions in the horizontal direction (spins defined on the vertices on the rough boundary is pinned to $\sigma_i = +1$) as shown in Fig.~\ref{suppfig:RBIM_coh-X}.

\begin{figure}
\centering
\begin{tikzpicture}[scale=1.0]
\definecolor{red1}{RGB}{240,83,90};
\definecolor{blue1}{RGB}{91,98,165};
\definecolor{myturquoise}{RGB}{83,195,189};
\definecolor{lightturquoise}{RGB}{64, 224, 208};
\definecolor{violet}{RGB}{207, 159, 255};
\foreach \x in {1,...,3,4}{
\draw[blue1!30,line width=1] (\x, 0.5) -- (\x, 3.5);
}
\foreach \y in {0.5,1.5,2.5,3.5}{
\draw[blue1!30,line width=1] (0,\y) -- (4.9,\y);
}
\foreach \x in {0.5,1.5,2.5,3.5,4.5}{
\foreach \y in {0.5,1.5,2.5,3.5}{
\filldraw[rounded corners=1pt, draw=black!80, fill=lightturquoise] (\x-0.1,\y-0.1) rectangle ++ (0.2,0.2);
}
}
\foreach \x in {1,2,3,4}{
\foreach \y in {1,2,3}{
\filldraw[rounded corners=1pt, draw=black!80, fill=lightturquoise] (\x-0.1,\y-0.1) rectangle ++ (0.2,0.2);
}
}
\foreach \x in {1,...,4}{
\foreach \y in {0.5,1.5,2.5,3.5}{
\filldraw[blue1] (\x,\y) circle (0.05);
}
}
\foreach \x in {-0.1,5}{
\foreach \y in {0.5,1.5,2.5,3.5}{
\draw[-latex, thick, blue1] (\x-0.1,\y-0.1) -- ++(0.3,0.3);
}
}
\draw[dotted,black!40,line width=0.5] (0.25,0.5) -- (0.25,3.5);
\foreach \y in {0.5,1.5,2.5,3.5}{
\draw[rounded corners=0.5pt, draw=black!40, fill=canaryyellow, rotate around={45:(0.25,\y)}] (0.25-0.075,\y-0.075) rectangle ++ (0.15,0.15);
}
\draw[dashed,black!60,line width=1.0] (-0.3,2.7) -- (5.3,2.7) -- (5.3,2.3) -- (-0.3,2.3) -- cycle;
\node[black] at (5.7,2.5) {$W_y$};
\draw[dashed,black!60,line width=1.0] (0.5,2.2) -- (4.5,2.2) -- (4.5,1.8) -- (0.5,1.8) -- cycle;
\node[black] at (5.7,2.0) {$V_y$};
\end{tikzpicture}
\caption{Tensor network representation of the partition function of RBIM associated with single-qubit unitary X-rotation. The yellow diamonds flip the random bond configuration $\eta_{ij}$, representing the defect insertion along a non-contractible path. Dashed boxed highlights the transfer matrices $W_y$ and $V_y$ in the vertical direction.}
\label{suppfig:RBIM_coh-X}
\end{figure}

The value of partition function $\calZ_{\sfa, s}$ and their ratio $\calZ_{1,s}/\calZ_{0,s}$ can be evaluated using the transfer matrix method as
\begin{align}
    \calZ_{\sfa,s} = \bra{+}_v^{\otimes L_x-1} W_{L_y+1} V_{L_y} W_{L_y} \cdots V_1 W_1\ket{+}_v^{\otimes L_x-1},
\end{align}
We present a tensor network representation of $\calZ_{\sfa,s}$ in Fig.~\ref{suppfig:RBIM_coh-X} with each local tensor is given by
\begin{align}
\begin{tikzpicture}[line cap=round,line join=round,x=1cm,y=1cm,baseline={(0,-\MathAxis pt)}]
\definecolor{blue1}{RGB}{91,98,165};
\definecolor{lightturquoise}{RGB}{64, 224, 208};
\draw[blue1!30,line width=1] (1, -0.5) node[below,blue1]{$\sigma_j$} -- (1, 0.5) node[above, blue1]{$\sigma_i$};
\filldraw[rounded corners=1pt, draw=black!80, fill=lightturquoise] (1-0.1,-0.1) rectangle ++ (0.2,0.2);
\end{tikzpicture} = \frac{1 + \eta_{ij}\sigma_i\sigma_j}{2}\cos\theta + \frac{1 - \eta_{ij}\sigma_i\sigma_j}{2}\ri \sin\theta. \label{eq:coh-X_tensor}
\end{align}
We note that the transfer matrix $W_y$ and $V_y$ in this model is generated by local Pauli-X and ZZ operator. Therefore, the transfer matrix contraction maps to a free fermion dynamics in 1+1d.

After applying the recovery string in the most likely class, we recover a logical state
\begin{align}
\ket{\bar{\Psi}} = \frac{1}{\sqrt{1 + e^{-2\abs{\Delta F_s}}}} \left( \ket{\Psi_0} + e^{-\Delta F_s+\ri\phi}\ket{\Psi_1}\right).
\end{align}
We note that $\Delta F_s = \abs{\log\calZ_{1,s}/\calZ_{0,s}}$, and $e^{\ri\phi}$ is a $\mathrm{U}(1)$ phase as the partition function is generally complex for unitary X-rotation.

\subsection{X rotation and XX rotation on horizontal edges}\label{sec:coh-X_coh-XX}

We now derive the RBIM for the surface code under both single- and two-qubit Pauli-X rotations.
The two-qubit Pauli-X rotation breaks the Gaussianity of fermion representation in the transfer matrix contraction, therefore, allows a volume-law phase shown in the main text.
For simplicity, we consider the two-qubit Pauli-X rotation only occur on the neighboring horizontal edges.

The corrupted surface code under both single- and two-qubit X-rotation is given by
\begin{align}
\ket{\Psi} &= \prod_{\langle \ell, \ell' \rangle_h} e^{\ri \varphi X_\ell X_{\ell'}}\prod_\ell e^{\ri \theta X_\ell} \ket{\Psi_0} = \sum_\calC A(\calC) X^\calC \ket{\Psi_0}.
\end{align}

In the ML decoding, for a given syndrome $s$, one can divide the error strings into two homological classes and evaluate the total probability of strings in each class
\begin{align}
    P_{\sfa,s} = \left|\bra{\Psi_0} X^{\calR_\sfa} \hat{\Pi}_s \ket{\Psi}\right|^2.
\end{align}
where $\calR_\sfa$ with $\sfa = 0,1$ is a reference string of class $\sfa$.

\begin{figure}
\centering
\begin{tikzpicture}[scale=1.0]
\definecolor{red1}{RGB}{240,83,90};
\definecolor{blue1}{RGB}{91,98,165};
\definecolor{myturquoise}{RGB}{83,195,189};
\definecolor{lightturquoise}{RGB}{64, 224, 208};
\definecolor{violet}{RGB}{207, 159, 255};
\foreach \x in {1,...,3,4}{
\draw[blue1!30,line width=1] (\x, 0.5) -- (\x, 3.5);
}
\foreach \y in {0.5,1.5,2.5,3.5}{
\draw[blue1!30,line width=1] (0,\y) -- (4.9,\y);
}
\foreach \x in {0.5,1.5,2.5,3.5,4.5}{
\foreach \y in {0.5,1.5,2.5,3.5}{
\filldraw[rounded corners=1pt, draw=black!80, fill=lightturquoise] (\x-0.1,\y-0.1) rectangle ++ (0.2,0.2);
}
}
\foreach \x in {1,2,3,4}{
\foreach \y in {1,2,3}{
\filldraw[rounded corners=1pt, draw=black!80, fill=lightturquoise] (\x-0.1,\y-0.1) rectangle ++ (0.2,0.2);
}
}
\foreach \x in {1,...,4}{
\foreach \y in {0.5,1.5,2.5,3.5}{
\filldraw[blue1] (\x,\y) circle (0.05);
\draw[violet, line width=1pt] (\x+0.25,\y) arc[start angle=-30, end angle=-150, radius=0.28867513459];
\filldraw[rounded corners=1pt, draw=black!80, fill=violet] (\x-0.25-0.075,\y-0.075) rectangle ++ (0.15,0.15);
\filldraw[rounded corners=1pt, draw=black!80, fill=violet] (\x+0.25-0.075,\y-0.075) rectangle ++ (0.15,0.15);
}
}
\foreach \x in {-0.1,5}{
\foreach \y in {0.5,1.5,2.5,3.5}{
\draw[-latex, thick, blue1] (\x-0.1,\y-0.1) -- ++(0.3,0.3);
}
}
\end{tikzpicture}
\caption{Tensor network representation of the partition function of RBIM associated with single-qubit unitary X-rotation on every qubit and two-qubit unitary XX-rotation on horizontal edges. The teal and purple boxes represent the tensor associated with single- and two-qubit Pauli-X rotations, respectively. The blue vertex represents a delta tensor. The summation over the contracted legs accounts for the summation over the Ising spin configurations in the partition function.}
\label{suppfig:RBIM_coh-X_coh-XX}
\end{figure}

The probability of each decoding class can be again formulated as the partition function of the RBIM.
We introduce the variables $\eta_{ij}$ associated with each bond to denote the reference string $\calR_\sfa$ and Ising spins variables $\sigma_i$ on each vertex to represent the error string $\calC_\sfa$,
\begin{align}
    \calR_{\sfa,ij} = \frac{1 - \eta_{ij}}{2}, \quad \calC_{\sfa,ij} = \frac{1 - \eta_{ij}\sigma_i \sigma_j}{2}.
\end{align}
The partition function $\calZ_\sfa = \sum_{\{\sigma_i\}} e^{-H(\{\sigma_i\})}$ describes the amplitude of all error strings in the class $\sfa$ has a tensor network representation in Fig.~\ref{suppfig:RBIM_coh-X_coh-XX}.
The partition function can be written in terms of the transfer matrix
\begin{align}
    \calZ_{\sfa,s} = \bra{+}_v^{\otimes L_x-1} W_{L_y+1} V_{L_y} W_{L_y} \cdots V_1 W_1 \ket{+}_v^{\otimes L_x-1}.
\end{align}
The transfer matrix associated with the vertical bonds takes the form
\begin{align}
    V_y = \begin{tikzpicture}[line cap=round,line join=round,x=1cm,y=1cm,baseline={(0,-\MathAxis pt)}]
\definecolor{blue1}{RGB}{91,98,165};
\definecolor{lightturquoise}{RGB}{64, 224, 208};
\foreach \x in {1,...,3,4}{
\draw[blue1!30,line width=1] (\x, -0.5) -- (\x, 0.5);
}
\foreach \x in {1,2,3,4}{
\foreach \y in {0}{
\filldraw[rounded corners=1pt, draw=black!80, fill=lightturquoise] (\x-0.1,\y-0.1) rectangle ++ (0.2,0.2);
}
}
\end{tikzpicture}
\end{align}
where each local tensor is given by
\begin{align}
\begin{tikzpicture}[line cap=round,line join=round,x=1cm,y=1cm,baseline={(0,-\MathAxis pt)}]
\definecolor{blue1}{RGB}{91,98,165};
\definecolor{lightturquoise}{RGB}{64, 224, 208};
\draw[blue1!30,line width=1] (1, -0.5) node[below,blue1]{$\sigma_j$} -- (1, 0.5) node[above, blue1]{$\sigma_i$};
\filldraw[rounded corners=1pt, draw=black!80, fill=lightturquoise] (1-0.1,-0.1) rectangle ++ (0.2,0.2);
\end{tikzpicture} = \frac{1 + \eta_{ij}\sigma_i\sigma_j}{2}\cos\theta + \frac{1 - \eta_{ij}\sigma_i\sigma_j}{2}\ri \sin\theta.
\end{align}
The transfer matrix associated with the vertical bond takes the form
\begin{align}
W_y = \begin{tikzpicture}[line cap=round,line join=round,x=1cm,y=1cm,baseline={(0,-\MathAxis pt)}]
\definecolor{blue1}{RGB}{91,98,165};
\definecolor{lightturquoise}{RGB}{64, 224, 208};
\definecolor{violet}{RGB}{207, 159, 255};
\foreach \x in {1,...,3,4}{
\draw[blue1!30,line width=1] (\x, -0.5) -- (\x, 0.5);
}
\draw[blue1!30,line width=1] (0,0) -- (5, 0);
\foreach \x in {1,...,3,4}{
\draw[violet, line width=1pt] (\x+0.25,0) arc[start angle=-30, end angle=-150, radius=0.28867513459];
\filldraw[rounded corners=1pt, draw=black!80, fill=violet] (\x-0.25-0.075,-0.075) rectangle ++ (0.15,0.15);
\filldraw[rounded corners=1pt, draw=black!80, fill=violet] (\x+0.25-0.075,-0.075) rectangle ++ (0.15,0.15);
}
\foreach \x in {0.5,1.5,2.5,3.5,4.5}{
\filldraw[rounded corners=1pt, draw=black!80, fill=lightturquoise] (\x-0.1,-0.1) rectangle ++ (0.2,0.2);
}
\foreach \x in {1,...,4}{
\foreach \y in {0}{
\filldraw[blue1] (\x,\y) circle (0.05);
}
}
\draw[-latex, thick, blue1] (-0.1,-0.1) -- ++(0.3,0.3);
\draw[-latex, thick, blue1] (5.1,-0.1) -- ++(0.3,0.3);
\end{tikzpicture}
\end{align}
where each local tensor takes the form
\begin{align}
\begin{tikzpicture}[line cap=round,line join=round,x=1cm,y=1cm,baseline={(0,-\MathAxis pt)}]
\definecolor{blue1}{RGB}{91,98,165};
\definecolor{lightturquoise}{RGB}{64, 224, 208};
\draw[blue1!30,line width=1] (0.5, 0) node[left,blue1]{$\alpha_j$} -- (1.5, 0) node[right, blue1]{$\alpha_i$};
\filldraw[rounded corners=1pt, draw=black!80, fill=lightturquoise] (1-0.1,-0.1) rectangle ++ (0.2,0.2);
\end{tikzpicture} &= \frac{1 + \eta_{ij}\alpha_i\alpha_j}{2}\cos\theta + \frac{1 - \eta_{ij}\alpha_i\alpha_j}{2}\ri \sin\theta, \\
\begin{tikzpicture}[line cap=round,line join=round,x=1cm,y=1cm,baseline={(0,-\MathAxis pt)}]
\definecolor{red1}{RGB}{240,83,90};
\definecolor{blue1}{RGB}{91,98,165};
\definecolor{violet}{RGB}{207, 159, 255};
\draw[blue1!30,line width=1] (1, -0.5) node[below,color=blue1]{$\sigma_i$} -- (1, 0.5) node[above,color=blue1]{$\sigma_i$};
\draw[blue1!30,line width=1] (0.5,0) node[left,color=blue1]{$\alpha_i$} -- (1.5, 0) node[right,color=blue1]{$\alpha_i$};
\filldraw[blue1] (1,0) circle (0.05);
\draw[violet, line width=1pt] (1+0.25,0) arc[start angle=-30, end angle=-150, radius=0.28867513459];
\filldraw[rounded corners=1pt, draw=black!80, fill=violet] (1-0.25-0.075,-0.075) rectangle ++ (0.15,0.15);
\filldraw[rounded corners=1pt, draw=black!80, fill=violet] (1+0.25-0.075,-0.075) rectangle ++ (0.15,0.15);
\end{tikzpicture} &= \frac{1 + \sigma_i \alpha_i}{2} \cos\varphi + \frac{1 - \sigma_i \alpha_i}{2}\ri \sin\varphi.
\end{align}

Apply a recovery operator along a string in the most likely class, we recover a logical state
\begin{align}
    \ket{\bar{\Psi}} &= \frac{1}{\sqrt{P_s}} X^{\calR_\sfa} \hat{\Pi}_s \sum_\calC A(\calC) X^\calC \ket{\Psi_0} = \frac{1}{\sqrt{1 + e^{-2\abs{\Delta F_s}}}} \left(\ket{\Psi_0} + e^{-\Delta F_s+\ri\phi}\ket{\Psi_1}\right),
\end{align}
where $\Delta F_s = |\log(\calZ_{1,s}/\calZ_{0,s})|$ is the excess free energy of defect insertion.
We again note that the resulting logical state $\ket{\bar{\Psi}}$ is only a class function that does not depend on the choice $\calR_\sfa$ in class $\sfa$.

\subsection{X, Y, Z rotation and XX rotation on horizontal edges}\label{sec:coh-XYZ_coh-XX}
Here, we derive the statistical mechanics model for the surface code under general single-qubit rotation and two-qubit XX rotation on horizontal edges.
The corrupted state takes the form
\begin{align}
    \ket{\Psi} &= \prod_{\langle\ell,\ell'\rangle} e^{\ri \varphi X_\ell X_{\ell'}} \prod_\ell e^{\ri \theta \mathbf{n} \cdot \mathbf{S}_\ell} \ket{\Psi_0} = \prod_{\langle\ell,\ell'\rangle} (\cos\varphi + \ri \sin\varphi X_\ell X_{\ell'}) \prod_\ell (\cos\theta + \ri \sin\theta \mathbf{n} \cdot \mathbf{S}_\ell) \ket{\Psi_0} \nonumber \\
    &= \sum_{\calC_x, \calC_z} A(\calC_x, \calC_z) X^{\calC_x} Z^{\calC_z} \ket{\Psi_0},
\end{align}
where $\mathbf{n} = (n_x, n_y, n_z)$ is a normalized vector, and $\mathbf{n} \cdot \mathbf{S}_\ell = n_x X_\ell + n_y Y_\ell + n_z Z_\ell$ specify the axis of single-qubit rotation.

The ML decoding for the surface code under this error model has a key difference; the recovered logical state depends not only on the homological class of the recovery operator but also on which operator is chosen from the class.
Assuming the initial state is in the logical $Z$ eigenstate $\ket{\Psi_0}$, the logical state after correction is given by
\begin{align}
    \ket{\bar{\Psi}[\calR_x,\calR_z]} &= Z^{\calR_z} X^{\calR_x} \hat{\Pi}_s \ket{\Psi} = (-1)^{\text{\#crossing}(\calR_x, \calR_z)}\sum_{\calC_x,\calC_z} A(\calC_x, \calC_z)(-1)^{\text{\#crossing}(\calC_x, \calR_z)}  \bar{X}^{[\calC_x+\calR_x]}\bar{Z}^{[\calC_z+\calR_z]} \ket{\Psi_0} \nonumber \\
    &= (\calZ_{00} + \calZ_{01}) \ket{\Psi_0} + (\calZ_{10} + \calZ_{11}) \ket{\Psi_1}.
\end{align}
We note that the non-trivial sign factor for each error configuration $\calC_x$ that depends on the choice of the recovery string $\calR_z$. The overall factor $(-1)^{\text{\#crossing}(\calR_x, \calR_z)}$ can be neglected.
We express the post-correction logical state in terms of $\calZ_{\sfa\sfb}$, which is given by
\begin{align}
\calZ_{\sfa\sfb} = \sum_{\calC_x, \calC_z | [\calC_x] = \sfa, [\calC_z] = \sfb} A(\calC_x, \calC_z)(-1)^{\text{\#crossing}(\calC_x, \calR_z)}. \label{eq:pt_coh-XYZ_coh-XX}
\end{align}
Similarly, if the initial state is in the logical X eigenstate $\ket{\Psi_+}$, the logical state post-correction takes the form
\begin{align}
    \ket{\bar{\Psi}[\calR_x,\calR_z]} &= Z^{\calR_z} X^{\calR_x} \hat{\Pi}_s \ket{\Psi} = (-1)^{\text{\#crossing}(\calR_x, \calR_z)}\sum_{\calC_x,\calC_z} A(\calC_x, \calC_z)(-1)^{\text{\#crossing}(\calC_x, \calR_z)}  \bar{X}^{[\calC_x+\calR_x]}\bar{Z}^{[\calC_z+\calR_z]} \ket{\Psi_+} \nonumber \\
    &= (\calZ_{00} + \calZ_{10}) \ket{\Psi_+} + (\calZ_{01} - \calZ_{11}) \ket{\Psi_-}.
\end{align}
Here, the homology class is defined according to the choice of the recovery string $\calR_x$ and $\calR_z$, namely $[\calC_{x}] = \sfa$ if $[\calC_{x} + \calR_{x}] = \bar{X}^\sfa$, and similarly for Z string.

In practice, we choose a particular recovery scheme to remove the syndrome, dubbed \emph{straight gauge}.
Specifically, we choose the recovery strings that are only on the horizontal bonds in the square lattice and require $\calR_x$ and $\calR_z$ to terminate only on the top and the right boundary, respectively.
We compute the amplitudes $\calZ_{\sfa\sfb}$ for four classes ($\sfa, \sfb = 0,1$) and determine the $\calZ_{\sfa_0\sfb_0}$ with the largest magnitude.
In numerical simulation, we compute the defect free energy that detects the symmetry breaking in two copies of Ising spins
\begin{align}
    \Delta F_X = |\log\calZ_{\bar{\sfa}_0\sfb_0}/\calZ_{\sfa_0\sfb_0}|, \quad \Delta F_Z = |\log\calZ_{\sfa_0\bar{\sfb}_0}/\calZ_{\sfa_0\sfb_0}|.
\end{align}
where $\bar{\sfa}_0$ denote the homological class that is not $\sfa_0$.

\begin{figure}
\centering
\begin{tikzpicture}[scale=1.0]
\definecolor{red1}{RGB}{240,83,90};
\definecolor{blue1}{RGB}{91,98,165};
\definecolor{myturquoise}{RGB}{83,195,189};
\definecolor{lightturquoise}{RGB}{64, 224, 208};
\definecolor{canaryyellow}{RGB}{255, 255, 143};
\foreach \x in {1,...,3,4}{
\draw[blue1!30,line width=1] (\x, 0.5) -- (\x, 3.5);
}
\foreach \y in {0.5,1.5,2.5,3.5}{
\draw[blue1!30,line width=1] (0,\y) -- (4.9,\y);
}
\foreach \x in {0.5,1.5,...,3.5,4.5}{
\draw[red1!30,line width=1] (\x,0) -- (\x, 3.9);
}
\foreach \y in {1,2,3}{
\draw[red1!30,line width=1] (0.5,\y) -- (4.5, \y);
}
\foreach \x in {0.5,1.5,2.5,3.5,4.5}{
\foreach \y in {0.5,1.5,2.5,3.5}{
\filldraw[rounded corners=1pt, draw=black!80, fill=lightturquoise] (\x-0.1,\y-0.1) rectangle ++ (0.2,0.2);
\filldraw[pattern=north east lines, draw=none] (\x-0.1,\y-0.1) rectangle ++ (0.2,0.2);
\draw[mint, line width=1pt, dashed] (\x+0.35,\y) arc[start angle=30, end angle=150, radius=0.40414518843];
\filldraw[draw=black!80, fill=mint] (\x+0.35,\y) circle (0.05);
\filldraw[draw=black!80, fill=mint] (\x-0.35,\y) circle (0.05);
}
}
\foreach \x in {1,2,3,4}{
\foreach \y in {1,2,3}{
\filldraw[rounded corners=1pt, draw=black!80, fill=lightturquoise] (\x-0.1,\y-0.1) rectangle ++ (0.2,0.2);
\filldraw[pattern=north east lines, draw=none] (\x-0.1,\y-0.1) rectangle ++ (0.2,0.2);
}
}
\foreach \x in {0.5,1.5,...,3.5,4.5}{
\foreach \y in {1,...,3}{
\draw[thick,red1,fill=white] (\x,\y) circle (0.05);
}
}
\foreach \x in {0.5,1.5,...,3.5,4.5}{
\foreach \y in {-0.1,4}{
\draw[-latex, thick, red1] (\x-0.1,\y-0.1) -- ++(0.3,0.3);
}
}
\foreach \x in {1,...,4}{
\foreach \y in {0.5,1.5,2.5,3.5}{
\filldraw[blue1] (\x,\y) circle (0.05);
\draw[violet, line width=1pt] (\x+0.3,\y) arc[start angle=-30, end angle=-150, radius=0.34641016151];
\filldraw[rounded corners=0.5pt, draw=black!80, fill=violet] (\x-0.3-0.05,\y-0.05) rectangle ++ (0.1,0.1);
\filldraw[rounded corners=0.5pt, draw=black!80, fill=violet] (\x+0.3-0.05,\y-0.05) rectangle ++ (0.1,0.1);
}
}
\foreach \x in {-0.1,5}{
\foreach \y in {0.5,1.5,2.5,3.5}{
\draw[-latex, thick, blue1] (\x-0.1,\y-0.1) -- ++(0.3,0.3);
}
}
\end{tikzpicture}
\caption{Tensor network representation of the complex weight partition function associated with general single-qubit rotation and two-qubit Pauli-X rotations. The model involves two copies of Ising spins associated with X and Z error strings on the blue and red vertices, respectively. The teal and purple tensors are governed by the single-qubit and two-qubit rotations. The green tensors account for the sign factor related to the crossing between $\calC_x$ and $\calR_z$.}
\label{suppfig:RBIM_coh-XYZ_coh-XX}
\end{figure}

The amplitude $\calZ_{\sfa\sfb}$ in each homological class maps to the partition function of the complex weight stat-mech model. 
We use binary variables $\eta_{ij}$ and $\xi_{\bi\bj}$ associated with each link $\ell = \langle i, j\rangle = \langle \bi, \bj\rangle $ to represent the recovery string, $\zeta_{x,ij}$ and $\zeta_{z,\bi\bj}$ to denote the logical string, and further introduce Ising spins $\sigma_i$ and $\tau_\bi$ on the vertices $i$ and $\bi$ of the direct and the dual lattice, respectively, to represent the error configuration
\begin{equation}
\begin{aligned}
\calR_{x,ij} = \frac{1 - \eta_{ij}}{2}, \quad \calR_{z,\bi\bj} = \frac{1 - \xi_{\bi\bj}}{2},& \quad \bar{X}_{ij} = \frac{1 - \zeta_{x,ij}}{2}, \quad \bar{Z}_{\bi\bj} = \frac{1 - \zeta_{z,\bi\bj}}{2}, \\
\calC_{x,ij} = \frac{1 - \eta_{ij}\zeta_{x,ij}\sigma_i \sigma_j}{2},& \quad \calC_{z,\bi\bj} = \frac{1 - \xi_{\bi\bj}\zeta_{z,\bi\bj}\tau_\bi \tau_\bj}{2}.
\end{aligned}
\end{equation}
In class $\sfa,\sfb \neq 1$, $\zeta_{x(z),\ell}$ takes $-1$ value along a non-contractible path, such that $\calC_{x(z)}$ denotes a string that is homologically inequvalent to $\calR_{x(z)}$. 
We note the logical string does not contribute the sign factor in Eq.~\ref{eq:pt_coh-XYZ_coh-XX} and can be chosen arbitrarily.

The partition function takes a tensor network representation as in Fig.~\ref{suppfig:RBIM_coh-XYZ_coh-XX} and can be written in terms of the transfer matrix as
\begin{align}
    \calZ_{\sfa\sfb}[\calR_x, \calR_z] = \bra{\uparrow}_p^{\otimes L_x} \bra{+}_v^{\otimes L_x-1} W_{L_y+1} V_{L_y} W_{L_y} \cdots V_1 W_1 \ket{\uparrow}_p^{\otimes L_x} \ket{+}_v^{\otimes L_x-1}
\end{align}
where $W_y$ and $V_y$ are the transfer matrix associated with the horizontal and the vertical bonds on the direct lattice, respectively.

The transfer matrices associated with horizontal bonds take the following MPO representation
\begin{align}
W_y = \begin{tikzpicture}[line cap=round,line join=round,x=1cm,y=1cm,baseline={(0,-\MathAxis pt)}]
\foreach \x in {1,...,3,4}{
\draw[blue1!30,line width=1] (\x, -0.5) -- (\x, 0.5);
}
\foreach \x in {0.5,1.5,...,3.5,4.5}{
\draw[red1!30,line width=1] (\x,-0.5) -- (\x, 0.5);
}
\foreach \y in {0}{
\draw[blue1!30,line width=1] (0,\y) -- (5, \y);
}
\foreach \x in {1,...,3,4}{
\draw[violet, line width=1pt] (\x+0.25,0) arc[start angle=-30, end angle=-150, radius=0.28867513459];
\filldraw[rounded corners=1pt, draw=black!80, fill=violet] (\x-0.25-0.075,-0.075) rectangle ++ (0.15,0.15);
\filldraw[rounded corners=1pt, draw=black!80, fill=violet] (\x+0.25-0.075,-0.075) rectangle ++ (0.15,0.15);
}
\foreach \x in {0.5,1.5,2.5,3.5,4.5}{
\foreach \y in {0}{
\filldraw[rounded corners=1pt, draw=black!80, fill=lightturquoise] (\x-0.1,\y-0.1) rectangle ++ (0.2,0.2);
\filldraw[pattern=north east lines, draw=none] (\x-0.1,\y-0.1) rectangle ++ (0.2,0.2);
}
}
\foreach \x in {1,...,4}{
\foreach \y in {0}{
\filldraw[blue1] (\x,\y) circle (0.05);
}
}
\draw[-latex, thick, blue1] (-0.1,-0.1) -- ++(0.3,0.3);
\draw[-latex, thick, blue1] (5.1,-0.1) -- ++(0.3,0.3);
\foreach \x in {1,3}{
\draw[mint, line width=1pt, dashed] (\x,0.2) -- (\x+1,0.2);
\filldraw[draw=black!80, fill=mint] (\x,0.2) circle (0.05);
\filldraw[draw=black!80, fill=mint] (\x+1,0.2) circle (0.05);
}
\foreach \x in {2}{
\draw[mint, line width=1pt, dashed] (\x,0.4) -- (\x+1,0.4);
\filldraw[draw=black!80, fill=mint] (\x,0.4) circle (0.05);
\filldraw[draw=black!80, fill=mint] (\x+1,0.4) circle (0.05);
}
\draw[mint, line width=1pt, dashed] (0,0.4) -- (1,0.4);
\filldraw[draw=black!80, fill=mint] (1,0.4) circle (0.05);
\draw[mint, line width=1pt, dashed] (4,0.4) -- (5,0.4);
\filldraw[draw=black!80, fill=mint] (4,0.4) circle (0.05);
\end{tikzpicture}
\end{align}
where each local tensor is given by
\begin{align}
\begin{tikzpicture}[line cap=round,line join=round,x=1cm,y=1cm,baseline={(0,-\MathAxis pt)}]
\definecolor{red1}{RGB}{240,83,90};
\definecolor{blue1}{RGB}{91,98,165};
\definecolor{myturquoise}{RGB}{83,195,189};
\definecolor{lightturquoise}{RGB}{64, 224, 208};
\definecolor{canaryyellow}{RGB}{255, 255, 143};
\draw[red1!30,line width=1] (1, -0.5) node[below,color=red1]{$\tau_\bj$} -- (1, 0.5) node[above,color=red1]{$\tau_\bi$};
\draw[blue1!30,line width=1] (0.5,0) node[left,color=blue1]{$\alpha_i$} -- (1.5, 0) node[right,color=blue1]{$\alpha_j$};
\filldraw[rounded corners=1pt, draw=black!80, fill=lightturquoise] (1-0.1,-0.1) rectangle ++ (0.2,0.2);
\filldraw[pattern=north east lines, draw=none] (1-0.1,-0.1) rectangle ++ (0.2,0.2);
\end{tikzpicture}
= &\frac{1 + \eta_{ij}\zeta_{x,ij}\alpha_i\alpha_j}{2}\frac{1 + \xi_{\bi\bj}\zeta_{z,\bi\bj}\tau_\bi \tau_\bj}{2} \cos\theta + \frac{1 - \eta_{ij}\zeta_{x,ij}\alpha_i\alpha_j}{2}\frac{1 + \xi_{\bi\bj}\zeta_{z,\bi\bj}\tau_\bi \tau_\bj}{2} \ri n_x \sin\theta\nonumber \\
&- \frac{1 - \eta_{ij}\zeta_{x,ij}\alpha_i\alpha_j}{2}\frac{1 - \xi_{\bi\bj}\zeta_{z,\bi\bj}\tau_\bi \tau_\bj}{2} n_y\sin\theta + \frac{1 + \eta_{ij}\zeta_{x,ij}\alpha_i\alpha_j}{2}\frac{1 - \xi_{\bi\bj}\zeta_{z,\bi\bj}\tau_\bi \tau_\bj}{2} \ri n_z \sin\theta, \\
\begin{tikzpicture}[line cap=round,line join=round,x=1cm,y=1cm,baseline={(0,-\MathAxis pt)}]
\definecolor{red1}{RGB}{240,83,90};
\definecolor{blue1}{RGB}{91,98,165};
\definecolor{violet}{RGB}{207, 159, 255};
\draw[blue1!30,line width=1] (1, -0.5) node[below,color=blue1]{$\sigma_i$} -- (1, 0.5) node[above,color=blue1]{$\sigma_i$};
\draw[blue1!30,line width=1] (0.5,0) node[left,color=blue1]{$\alpha_i$} -- (1.5, 0) node[right,color=blue1]{$\alpha_i$};
\filldraw[blue1] (1,0) circle (0.05);
\draw[violet, line width=1pt] (1+0.25,0) arc[start angle=-30, end angle=-150, radius=0.28867513459];
\filldraw[rounded corners=1pt, draw=black!80, fill=violet] (1-0.25-0.075,-0.075) rectangle ++ (0.15,0.15);
\filldraw[rounded corners=1pt, draw=black!80, fill=violet] (1+0.25-0.075,-0.075) rectangle ++ (0.15,0.15);
\end{tikzpicture} = & \frac{1 + \sigma_i \alpha_i}{2} \cos\varphi + \frac{1 - \sigma_i \alpha_i}{2}\ri \sin\varphi, \\
\begin{tikzpicture}[line cap=round,line join=round,x=1cm,y=1cm,baseline={(0,-\MathAxis pt)}]
\definecolor{red1}{RGB}{240,83,90};
\definecolor{blue1}{RGB}{91,98,165};
\definecolor{violet}{RGB}{207, 159, 255};
\definecolor{mint}{RGB}{152, 251, 152};
\draw[blue1!30,line width=1] (0, -0.2) node[below,color=blue1]{$\sigma_i$} -- (0, 0.2) node[above,color=blue1]{$\sigma_i$};
\draw[blue1!30,line width=1] (1, -0.2) node[below,color=blue1]{$\sigma_j$} -- (1, 0.2) node[above,color=blue1]{$\sigma_j$};
\draw[mint, line width=1pt, dashed] (0,0) -- (1,0);
\filldraw[draw=black!80, fill=mint] (0,0) circle (0.05);
\filldraw[draw=black!80, fill=mint] (1,0) circle (0.05);
\end{tikzpicture}
 = & (-1)^{\frac{1-\xi_{\bi\bj}}{2}\frac{1 - \eta_{ij}\zeta_{x,ij}\sigma_i\sigma_j}{2}}.
\end{align}

The transfer matrix associated with vertical bonds takes the form
\begin{align}
V_y =
\begin{tikzpicture}[line cap=round,line join=round,x=1cm,y=1cm,baseline={(0,-\MathAxis pt)}]
\definecolor{red1}{RGB}{240,83,90};
\definecolor{blue1}{RGB}{91,98,165};
\definecolor{myturquoise}{RGB}{83,195,189};
\definecolor{lightturquoise}{RGB}{64, 224, 208};
\definecolor{canaryyellow}{RGB}{255, 255, 143};
\definecolor{mint}{RGB}{152, 251, 152};
\foreach \x in {1,...,3,4}{
\draw[blue1!30,line width=1] (\x, -0.5) -- (\x, 0.5);
}
\foreach \x in {0.5,1.5,...,3.5,4.5}{
\draw[red1!30,line width=1] (\x,-0.5) -- (\x, 0.5);
}
\foreach \y in {0}{
\draw[red1!30,line width=1] (0.5,\y) -- (4.5, \y);
}
\foreach \x in {1,2,3,4}{
\filldraw[rounded corners=1pt, draw=black!80, fill=lightturquoise] (\x-0.1,-0.1) rectangle ++ (0.2,0.2);
\filldraw[pattern=north east lines, draw=none] (\x-0.1,-0.1) rectangle ++ (0.2,0.2);
\draw[mint, line width=1pt, dashed] (\x,0.3) arc[start angle=60, end angle=-60, radius=0.34641016151];
\filldraw[draw=black!80, fill=mint] (\x,0.3) circle (0.05);
\filldraw[draw=black!80, fill=mint] (\x,-0.3) circle (0.05);
}
\foreach \x in {0.5,1.5,...,3.5,4.5}{
\foreach \y in {0}{
\draw[thick,red1,fill=white] (\x,\y) circle (0.05);
}
}
\end{tikzpicture}
\end{align}
where each local tensor is given by
\begin{align}
\begin{tikzpicture}[line cap=round,line join=round,x=1cm,y=1cm,baseline={(0,-\MathAxis pt)}]
\definecolor{red1}{RGB}{240,83,90};
\definecolor{blue1}{RGB}{91,98,165};
\definecolor{lightturquoise}{RGB}{64, 224, 208};
\definecolor{mint}{RGB}{152, 251, 152};
\draw[blue1!30,line width=1] (1, -0.5) node[below,color=blue1]{$\sigma_j$} -- (1, 0.5) node[above,color=blue1]{$\sigma_i$};
\draw[red1!30,line width=1] (0.5,0) node[left,color=red1]{$\tau_\bi$} -- (1.5, 0) node[right,color=red1]{$\tau_\bj$};
\filldraw[rounded corners=1pt, draw=black!80, fill=lightturquoise] (1-0.1,-0.1) rectangle ++ (0.2,0.2);
\filldraw[pattern=north east lines, draw=none] (1-0.1,-0.1) rectangle ++ (0.2,0.2);
\draw[mint, line width=1pt, dashed] (1,0.3) arc[start angle=60, end angle=-60, radius=0.34641016151];
\filldraw[draw=black!80, fill=mint] (1,0.3) circle (0.05);
\filldraw[draw=black!80, fill=mint] (1,-0.3) circle (0.05);
\end{tikzpicture}
= &\frac{1 + \eta_{ij}\zeta_{x,ij}\sigma_i\sigma_j}{2}\frac{1 + \xi_{\bi\bj}\zeta_{z,\bi\bj}\tau_\bi \tau_\bj}{2} \cos\theta + \frac{1 - \eta_{ij}\zeta_{x,ij}\sigma_i\sigma_j}{2}\frac{1 + \xi_{\bi\bj}\zeta_{z,\bi\bj}\tau_\bi \tau_\bj}{2} \ri n_x \sin\theta (-1)^{\frac{1 -\xi_{\bi\bj}}{2}}\nonumber \\
&- \frac{1 - \eta_{ij}\zeta_{x,ij}\sigma_i\sigma_j}{2}\frac{1 - \xi_{\bi\bj}\zeta_{z,\bi\bj}\tau_\bi \tau_\bj}{2} n_y\sin\theta (-1)^{\frac{1 -\xi_{\bi\bj}}{2}} + \frac{1 + \eta_{ij}\zeta_{x,ij}\sigma_i\sigma_j}{2}\frac{1 - \xi_{\bi\bj}\zeta_{z,\bi\bj}\tau_\bi \tau_\bj}{2} \ri n_z \sin\theta.
\end{align}

\section{Gauge invariance of entanglement in transfer matrix contraction}

In this section, we show that the entanglement in the transfer matrix contraction of the RBIM is a gauge invariant quantity that only depends on the observed syndrome but crucially does not depend on the choice of random bond configuration.

In the RBIM for single- and two-qubit Pauli-X rotation, the gauge transformation is generated by flipping the four random bond variables $\eta_{ij}$ emanated from each blue vertex.
In the tensor network representation, the transformation is achieved by inserting Pauli-X operators (yellow diamond below) to the bond next to the tensors in teal, i.e.
\begin{align}
\begin{tikzpicture}[line cap=round,line join=round,x=1cm,y=1cm,baseline={(0,-\MathAxis pt)}]
\definecolor{red1}{RGB}{240,83,90};
\definecolor{blue1}{RGB}{91,98,165};
\definecolor{violet}{RGB}{207, 159, 255};
\draw[blue1!30,line width=1] (1, -1) -- (1, 1);
\draw[blue1!30,line width=1] (0.,0) -- (2, 0);
\filldraw[blue1] (1,0) circle (0.05);
\draw[violet, line width=1pt] (1+0.25,0) arc[start angle=-30, end angle=-150, radius=0.28867513459];
\filldraw[rounded corners=1pt, draw=black!80, fill=violet] (1-0.25-0.075,-0.075) rectangle ++ (0.15,0.15);
\filldraw[rounded corners=1pt, draw=black!80, fill=violet] (1+0.25-0.075,-0.075) rectangle ++ (0.15,0.15);
\filldraw[rounded corners=1pt, draw=black!80, fill=lightturquoise] (1-0.1,-0.8) rectangle ++ (0.2,0.2);
\filldraw[rounded corners=1pt, draw=black!80, fill=lightturquoise] (1-0.1,0.6) rectangle ++ (0.2,0.2);
\filldraw[rounded corners=1pt, draw=black!80, fill=lightturquoise] (1-0.7-0.1,-0.1) rectangle ++ (0.2,0.2);
\filldraw[rounded corners=1pt, draw=black!80, fill=lightturquoise] (1+0.6,-0.1) rectangle ++ (0.2,0.2);
\end{tikzpicture} \mapsto
\begin{tikzpicture}[line cap=round,line join=round,x=1cm,y=1cm,baseline={(0,-\MathAxis pt)}]
\definecolor{red1}{RGB}{240,83,90};
\definecolor{blue1}{RGB}{91,98,165};
\definecolor{violet}{RGB}{207, 159, 255};
\draw[blue1!30,line width=1] (1, -1) -- (1, 1);
\draw[blue1!30,line width=1] (0.,0) -- (2, 0);
\filldraw[blue1] (1,0) circle (0.05);
\draw[violet, line width=1pt] (1+0.25,0) arc[start angle=-30, end angle=-150, radius=0.28867513459];
\filldraw[rounded corners=1pt, draw=black!80, fill=violet] (1-0.25-0.075,-0.075) rectangle ++ (0.15,0.15);
\filldraw[rounded corners=1pt, draw=black!80, fill=violet] (1+0.25-0.075,-0.075) rectangle ++ (0.15,0.15);
\filldraw[rounded corners=1pt, draw=black!80, fill=lightturquoise] (1-0.1,-0.8) rectangle ++ (0.2,0.2);
\filldraw[rounded corners=1pt, draw=black!80, fill=lightturquoise] (1-0.1,0.6) rectangle ++ (0.2,0.2);
\filldraw[rounded corners=1pt, draw=black!80, fill=lightturquoise] (1-0.8-0.1,-0.1) rectangle ++ (0.2,0.2);
\filldraw[rounded corners=1pt, draw=black!80, fill=lightturquoise] (1+0.7,-0.1) rectangle ++ (0.2,0.2);
\filldraw[rounded corners=0.5pt, draw=black!80, fill=canaryyellow,rotate around={45:(1,0.4)}] (1-0.075,0.4-0.075) rectangle ++ (0.15,0.15);
\filldraw[rounded corners=0.5pt, draw=black!80, fill=canaryyellow,rotate around={45:(1,-0.4)}] (1-0.075,-0.4-0.075) rectangle ++ (0.15,0.15);
\filldraw[rounded corners=0.5pt, draw=black!80, fill=canaryyellow,rotate around={45:(1.5,0)}] (1.5-0.075,-0.075) rectangle ++ (0.15,0.15);
\filldraw[rounded corners=0.5pt, draw=black!80, fill=canaryyellow,rotate around={45:(0.5,0)}] (0.5-0.075,-0.075) rectangle ++ (0.15,0.15);
\end{tikzpicture} = 
\begin{tikzpicture}[line cap=round,line join=round,x=1cm,y=1cm,baseline={(0,-\MathAxis pt)}]
\definecolor{red1}{RGB}{240,83,90};
\definecolor{blue1}{RGB}{91,98,165};
\definecolor{violet}{RGB}{207, 159, 255};
\draw[blue1!30,line width=1] (1, -1) -- (1, 1);
\draw[blue1!30,line width=1] (0.,0) -- (2, 0);
\filldraw[blue1] (1,0) circle (0.05);
\draw[violet, line width=1pt] (1+0.5,0) arc[start angle=-30, end angle=-150, radius=0.57735026918];
\filldraw[rounded corners=1pt, draw=black!80, fill=violet] (1-0.5-0.075,-0.075) rectangle ++ (0.15,0.15);
\filldraw[rounded corners=1pt, draw=black!80, fill=violet] (1+0.5-0.075,-0.075) rectangle ++ (0.15,0.15);
\filldraw[rounded corners=1pt, draw=black!80, fill=lightturquoise] (1-0.1,-0.8) rectangle ++ (0.2,0.2);
\filldraw[rounded corners=1pt, draw=black!80, fill=lightturquoise] (1-0.1,0.6) rectangle ++ (0.2,0.2);
\filldraw[rounded corners=1pt, draw=black!80, fill=lightturquoise] (1-0.8-0.1,-0.1) rectangle ++ (0.2,0.2);
\filldraw[rounded corners=1pt, draw=black!80, fill=lightturquoise] (1+0.7,-0.1) rectangle ++ (0.2,0.2);
\filldraw[rounded corners=0.5pt, draw=black!80, fill=canaryyellow,rotate around={45:(1,0.4)}] (1-0.075,0.4-0.075) rectangle ++ (0.15,0.15);
\filldraw[rounded corners=0.5pt, draw=black!80, fill=canaryyellow,rotate around={45:(1,-0.4)}] (1-0.075,-0.4-0.075) rectangle ++ (0.15,0.15);
\filldraw[rounded corners=0.5pt, draw=black!80, fill=canaryyellow,rotate around={45:(1.2,0)}] (1.2-0.075,-0.075) rectangle ++ (0.15,0.15);
\filldraw[rounded corners=0.5pt, draw=black!80, fill=canaryyellow,rotate around={45:(0.8,0)}] (0.8-0.075,-0.075) rectangle ++ (0.15,0.15);
\end{tikzpicture}
\end{align}
The transformation keeps the tensor network invariant as four Pauli-X operators annihilate at the delta tensor at the blue vertex.

Importantly, the gauge transformation, while altering the transfer matrix, does not affect the entanglement in the transfer matrix contraction.
This is because, at any step during the transfer matrix contraction, the transformed wave function differs from the original wave function by a product of Pauli-X operators, which does not affect the entanglement.

In the stat-mech model for general single-qubit rotation and two-qubit Pauli-X rotation, we note that the gauge redundancy is associated with the choice of logical string $\zeta_{x,ij}$, $\zeta_{z,\bi\bj}$ (\textit{not} recovery string $\eta_{ij}$, $\xi_{\bi\bj}$). The gauge transformation similarly keeps the tensor network invariant and affects the wave function only by a product of Pauli-X operators at the intermediate step of the contraction. Therefore, the entanglement is invariant under the transformation.

\section{Isometric tensor network representation of the surface code}
\label{sec:isoTNS}

\begin{figure}
    \centering
    \includegraphics[]{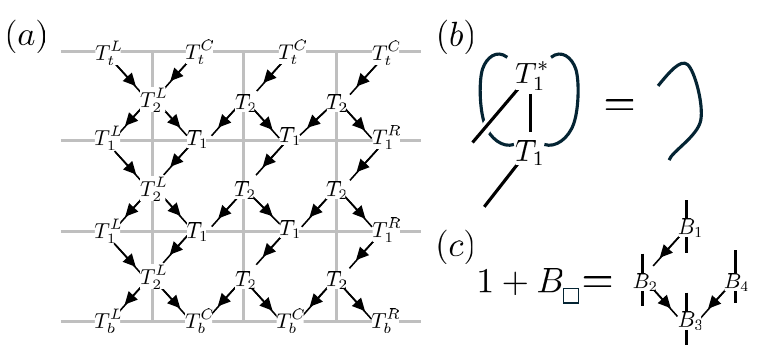}
    \caption{(a) Isometric tensor network representation for $4 \times 3$ surface cdoe. Each tensor is labeled by name and definitions of this tensor are in the text. Physical legs are supressed for clarity. (b) Orthogonality condition for isometric tensor; contracting over incoming legs yields an identity on outgoing legs. (c) Matrix product operator representation of the unnormalized plaquette projector}
    \label{suppfig:isoTNS}
\end{figure}

As a starting point for the syndrome sampling algorithm presented in Sec.~\ref{sec:sampling}, we need an isometric tensor network representation of the surface code.
Isometric tensor networks (isoTNS), a restriction of projected entangled-pair states in which each tensor is isometric, were originally introduced as a numerical ansatz for studying ground states of two-dimensional interacting quantum spin systems~\cite{zaletel2020isometric} but later have been used to numerically study time dynamics, fermions, and finite temperature properties~\cite{Lin_2022, dai2024fermionicisometrictensornetwork, Kadow_2023}.
Additionally, it has been shown that isoTNS can exactly represent string net fixed point wavefunctions, universal quantum computation, and critical states derived from classical statistical mechanics models~\cite{Soejima_2020,malz2024computationalcomplexityisometrictensor,liu2024simulating2dtopologicalquantum}.
The isoTNS for the surface code is shown graphically in Fig.~\ref{suppfig:isoTNS}(a), where the lattice is drawn with light grey lines, and the bonds between tensors are drawn in black.
To each leg of a tensor, we associate an arrow.
If we view the tensor as a matrix $A$ with outgoing legs grouped as the row index and incoming legs grouped as the column index, the matrix $A$ is isometric; i.e $A A^\dagger = \mathbb{I}$. 
Such constraints are commonly imposed on one-dimensional tensor networks, such as matrix product states, and are shown pictorially in Fig.~\ref{suppfig:isoTNS}(b)
As explained in Sec.~\ref{sec:sampling}, this isometric constraint enables an efficient syndrome sampling algorithm.

Now we describe how to construct an isoTNS for the surface code in the logical state $\ket{\Psi_+}$ consisting of $L_x \times L_y$ plaquettes.
Such a state will have $N = L_x(L_y+1) + (L_x-1)L_y = 2L_x L_y + L_x - L_y$ qubits.
We start with the state $\ket{+}^{\otimes N}$, which is automatically a $+1$ eigenstate of the $(L_x-1) L_y$ $A_s$ plaquette stabilizers.
Then on each of the $L_x L_y$ plaquettes, we apply the (unnormalized) projector $1 + B_\Box$ such that the resulting state is the surface code wavefunction $\ket{\Psi_+}$.
To see that we have this logical state, note that the horizontal $\overline{Z}$ logical and vertical $\overline{X}$ logical operators independently commute with the projector (while anticommuting with one another) and thus in expectation are entirely determined by the initial state; thus $\bra{\Psi_+} \overline{X} \ket{\Psi_+} \propto 1$ and $\bra{\Psi_+} \overline{Z} \ket{\Psi_+} \propto 0$.

To apply the projector in a way that produces the isoTNS with the isometric arrow structure shown in Fig.~\ref{suppfig:isoTNS}(a), we first write the projector $1 + B_\Box$ as a matrix product operator.
This projector acts on four spins in a plaquette and is diagonal in the computational basis.
It simply enforces that the 4-bit computational basis strings have even parity.
We write the projector, shown graphically in Fig.~\ref{suppfig:isoTNS}(c), as
\begin{align}
    \left(1 + B_\Box\right)^{\{\sigma\};\{\sigma'\}} = \sum_{\eta \gamma \chi} B^{[1]\sigma_1;\sigma_1'}_{\eta} B^{[2] \sigma_2; \sigma_2'}_{\eta \gamma} B^{[3] \sigma_3; \sigma_3'}_{\gamma \chi} B^{[4] \sigma_4; \sigma_4'}_{\chi},
\end{align}
where Greek letters indicate virtual bonds of the MPO, $\{\sigma\}$ are input physical indices, and $\{\sigma'\}$ are output physical indices.
Tensor $B^{[3]}$ is chosen to apply the parity constraint.
Tensors $B^{[1]}$ and $B^{[4]}$ are equivalent and simply copy the physical index into the virtual leg.
Tensor $B^{[2]}$ combines together (i.e. addition modulo 2, $\oplus$) the physical index and the virtual incoming $\eta$ from site $1$ into the outgoing virtual index $\gamma$.
Mathematically, these tensors are given by the following.
\begin{align} 
    B^{[1]\sigma;\sigma'}_{\eta} &= B^{[4]\sigma;\sigma'}_{\eta} = \delta_{\sigma, \sigma'} \delta_{\sigma, \eta} \\
    B^{[2] \sigma; \sigma'}_{\eta \gamma} &= \delta_{\sigma, \sigma'} \delta_{\gamma, \eta \oplus \sigma} \\
    B^{[3] \sigma; \sigma'}_{\gamma \chi} &= \delta_{\sigma, \sigma'} \delta_{\gamma \oplus \chi,\sigma}
\end{align}
The left and right rough boundaries require slightly modified MPOs to impose the projector on to $+1$ eigenstate of 3-body stabilizers $B_\sqsupset$ and $B_\sqsubset$, respectively.
Both MPOs simply apply the condition that the parity of bit strings on the three sites must be even and thus have the same form.
\begin{align}
    \left(1 + B_\sqsupset\right)^{\{\sigma\};\{\sigma'\}} &= \left(1 + B_\sqsubset\right)^{\{\sigma\};\{\sigma'\}} = \sum_{\eta \gamma} C^{[1]\sigma_1;\sigma_1'}_{\eta} C^{[2] \sigma_2; \sigma_2'}_{\eta \gamma} C^{[3] \sigma_3; \sigma_3'}_{\gamma} \\
    C^{[1]\sigma;\sigma'}_{\eta} &= \delta_{\sigma, \sigma'} \delta_{\sigma, \eta} \\
    C^{[2] \sigma; \sigma'}_{\eta \gamma} &= \delta_{\sigma, \sigma'} \delta_{\gamma, \eta \oplus \sigma} \\
    C^{[3] \sigma; \sigma'}_{\gamma} &= \delta_{\sigma, \sigma'} \delta_{\gamma,\sigma}
\end{align}

Now, we apply the appropriate MPO on each plaquette of the surface code to build up the correlated ground state.
Since the projectors commute with one another, the order in which they are applied is arbitrary.
Contracting with the $\ket{+}$ state is equivalent to summing over the physical index, so after we do this, we find the tensor network shown in Fig.~\ref{suppfig:isoTNS}(a), with 9 distinct types of tensors.
Our tensors each have 5 legs, which are labeled as $T^{p}_{u_L u_R, \ell_L \ell_R}$ for the physical $p$, upper left $u_L$, upper right $u_R$, lower left $\ell_L$, and lower right $\ell_R$ legs, respectively.
Now we can define the tensors that result from contracting the MPOs with the state $\ket{+}^{N}$.
Omitted indices on the tensor indicate a trivial bond dimension of 1.
Normalization factors have been introduced to ensure each tensor is properly isometric.
\begin{align}
    [T_t^L]^p_{\ell_R} &= \delta_{p,\ell_R} \\
    [T_t^C]^p_{\ell_L} &= \delta_{p, \ell_L} \\
    [T_1^L]^p_{u_R \ell_R} &= \delta_{p, u_R} \delta_{p, \ell_R} \\
    [T_1^R]^p_{u_L \ell_L} &= \delta_{p, u_L} \delta_{p, \ell_L} \\
    [T_1]^p_{u_L u_R \ell_L} &= \frac{1}{\sqrt{2}} \delta_{p, u_L + u_R} \delta_{p, \ell_L} \label{eq:T1} \\
    [T_2]^p_{u_R \ell_L \ell_R} &= \delta_{p, \ell_L} \delta_{p + u_R, \ell_R} \\
    [T_2^L]^p_{u_R \ell_L \ell_R} &= \frac{1}{\sqrt{2}} \delta_{u_L + p, \ell_L} \delta_{p + u_R, \ell_R} \\
    [T_b^L]^p_{u_R} &= \frac{1}{\sqrt{2}}\delta_{p,u_R} \\
    [T_b^R]^p_{u_L} &= \frac{1}{\sqrt{2}}\delta_{p,u_L} \\
    [T_b^C]^p_{u_R u_L} &= \frac{1}{2}\delta_{p,u_R + u_L}
\end{align}
All of these tensors are isometric from physical and upper legs to the lower legs.
As an example, consider $T_1$ given by Eq.~\eqref{eq:T1}. Let us show that it is isometric.
\begin{align*}
    \sum_{u_L u_R p} [T_1]^{p}_{u_L u_R \ell_L} \left([T_1]^{p}_{u_L u_R \ell_L'}\right)^* &= \frac{1}{2} \sum_{u_L u_R p} \delta_{p, u_L + u_R} \delta_{p, \ell_L} \delta_{p, u_L + u_R} \delta_{p, \ell_L'} \\
    &= \frac{1}{2} \sum_{u_L u_R} \delta_{\ell_L, u_L + u_R} \delta_{\ell_L, u_L + u_R} \delta_{\ell_L, \ell_L'} \\
    &= \frac{1}{2} \sum_{u_L u_R} \delta_{\ell_L, u_L + u_R} \delta_{\ell_L, \ell_L'} \\
    &= \frac{1}{2} [2 \delta_{\ell_L, 0} + 2 \delta_{\ell_L, 1}]\delta_{\ell_L, \ell_L'} \\
    &= \delta_{\ell_L, \ell_L'}
\end{align*}
Thus these tensors define an, at most, bond dimension $\chi=2$ isoTNS for the surface code with rough boundaries on left and right and smooth boundaries on the top and bottom.
This planar state encodes a single logical qubit with a vertical $X$ logical string and horizontal $Z$ logical string.

\begin{figure}
\centering
\begin{tikzpicture}[line cap=round,line join=round,x=1cm,y=1cm,baseline={(0, -\MathAxis pt)},scale=0.8]
\definecolor{mygreen}{RGB}{134, 175, 142};
\definecolor{mint}{RGB}{152, 251, 152};
\definecolor{violet}{RGB}{207, 159, 255};
\definecolor{mint2}{RGB}{218, 247, 166};
\definecolor{raspberry}{RGB}{227, 11, 92};
\foreach \x in {0,4,8,12}{
\draw[black!80, line width=1.0] (\x, 0) node[below] {$\ket{+}$} -- (\x, 3.5);
}
\foreach \x in {2,6,10}{
\draw[black!80, line width=1.0] (\x, 0) node[below] {$\ket{+}$} -- (\x, 3.5);
}

\foreach \x in {0,2,...,8,10}{
\filldraw[fill=mint, rounded corners=2pt, draw=black!80, line width=1.0] (\x-0.2,0.25*\x+0.15) rectangle ++(2.4,0.2);
}
\foreach \x in {0,2,...,10,12}{
\filldraw[fill=violet,rounded corners=2pt, draw=black!80, line width=1.0] (\x-0.2,3.3) rectangle (\x+0.2,3.0);
}
\fill[raspberry, opacity=0.3] (-0.2,0.1) -- (4.5,0.1) -- (2.2,3.4) -- (-0.2,3.4) -- cycle;
\draw[raspberry, rounded corners, line width=1.5] (-0.2, 3.55) -- (2.2,3.55) -- (2.2, 4.45) -- (-0.2,4.45) -- cycle;
\draw[raspberry, line width=1.5] (1.0, 3.75) -- (1.4, 4.25);
\draw[raspberry, line width=1.5] (.55, 3.85) arc[start angle=150, end angle=30, radius=0.5];
\node[raspberry] at (-1.0,3.8) {$\mathcal{O}_{1,2}$};
\end{tikzpicture}
\caption{Sampling two-qubit observables from a one-dimensional quantum state, represented by a sequential unitary state. The green gates are sequantial unitary gates while the purple gates represent the unitary noise. As discussed in the text, the sequential structure allows for efficient sampling.}
\label{suppfig:sequential}
\end{figure}

\section{Algorithm for syndrome sampling}
\label{sec:sampling}

Given the isoTNS for the unperturbed surface code logical $\ket{+}$ state, defined in Sec.~\ref{sec:isoTNS}, we want to apply errors and sample syndromes.
An in-depth description of our sampling algorithm and extensions to generic errors described by arbitrary Kraus operators will appear in forthcoming work. Here we only work with unitary errors such that the corrupted state remains pure.

To begin, consider the one-dimensional state on 7 qubits shown in Fig.~\ref{suppfig:sequential}.
This uncorrupted state $\ket{\Psi_0}$ is prepared by a sequential quantum circuit, shown here as a staircase unitary circuit.
Following the state preparation circuit, shown in green, we apply single qubit unitary noise, denoted by purple gates, yielding the corrupted state $\ket{\Psi}$.
Suppose our goal is to measure some two-qubit, mutually-commuting Pauli operator on each neighboring pair of operators, say $\mathcal{O}_{i,i+1} = Z_iZ_{i+1}$.
The probability distribution of outcomes is given by
\begin{align}
    P(+1) = \bra{\Psi} \frac{1 + \mathcal{O}}{2} \ket{\psi}, \:\:\: P(-1) = \bra{\Psi} \frac{1 - \mathcal{O}}{2} \ket{\psi}.
\end{align}
If the state $\ket{\Psi}$ were represented by a matrix product state (MPS), we could simply contract the network corresponding to these operator expectation values to find the measurement distribution.
However, from this sequential circuit perspective, we see that network contraction is unnecessary.
All we need is the reduced density matrix (RDM) on the first two sites, denoted as $\rho_{1,2} = \mathrm{Tr}_{3,\ldots,7} \ket{\Psi} \bra{\Psi}$.
This only requires the first three qubits and first two state preparation unitaries; the rest of the qubits and unitaries do not contribute, as the unitary gates and noise cancel with their conjugate when tracing the density matrix.
Note that this would not happen if we had non-unitary noise.
Thus in our case of unitary noise, we can efficiently find the RDM for the first two qubits as
\begin{align}
    \rho_{1,2} = \mathrm{Tr}_3 \bigg[ V_3 V_2 V_1 U_{23} U_{12} \ket{+++}\bra{+++} U_{12}^\dagger U_{23}^\dagger V_1^\dagger V_2^\dagger V_3^\dagger \bigg]
\end{align}
After doing so, we can measure the operator $\mathcal{O}_{1,2}$ and project $\ket{\psi}$ onto the eigenspace of the measurement outcome, which due to multi-qubit measurement does a priori not collapse any qubits onto a definite state.
Then, note that qubit 1 is not required in any future measurements, so we can remove it from the system by projecting it onto a definite state or measuring it in an arbitrary single qubit basis.
This yields a six qubit system.
We then proceed to measuring $\mathcal{O}_{2,3}$ by finding the RDM on sites 2 and 3, which again is simplified due to the sequential circuit structure and unitary noise.
This process is repeated until all two-qubit gates are measured.

To extend this to 2D isoTNS and the surface code, first note that the unitary noise we apply retains the isometric property of the network.
On each row, we apply unitary rotation errors, and in the most general case we consider, the noise corrupted state can be written as
\begin{align}
    \ket{\Psi} &= \prod_{\langle\ell,\ell'\rangle} e^{\ri \varphi X_\ell X_{\ell'}} \prod_\ell e^{\ri \theta \mathbf{n} \cdot \mathbf{S}_\ell} \ket{\Psi_0}.
\end{align}
The single qubit errors can be applied on every tensor without affecting the bond dimension or the isometric conditions on each tensor.
The two-qubit error on primal rows is contained entirely within each row and thus can be viewed analogously to the single qubit errors in Fig.~\ref{suppfig:sequential}.
Next, note that any isometric tensor network can be viewed as a state preparation sequential quantum circuit~\cite{Sch_n_2005, Ba_uls_2008, Wei_2022}.
The isometric tensors can be viewed into unitary quantum gates acting on input qubits in a definite state, and the order of their application is \textit{against} the flow of isometric arrows~\cite{Anand_2023}.

With this, the algorithm for sampling stabilizers in the unitary error corrupted surface code is analogous to the procedure described above for a 1D state.
Starting on the bottom most row, we repeat the following, moving upwards, until all syndromes have been measured:
\begin{enumerate}
    \item Sample the stabilizer contained on either two (boundary) or three (bulk) rows by finding the RDM for the involved rows. Due to the isometric structure of the network, this RDM can be found efficiently as rows above the ones we sample trace to the identity.
    \item Project $\ket{\Psi}$ into the eigenspace of the stabilizer with the desired measurement outcome. We apply the projectors within the rows being sampled and represent the result as an MPS.
    \item Measure out qubits in the computational basis after all stabilizers in which they are involved are measured. This ensures that the number of qubits in the MPS does now grow as one continues to sample
\end{enumerate}
As we proceed row-by-row upwards sampling syndromes, the bond dimension of the MPS representing the projected rows will grow and require truncation.
Thus, this algorithm is only approximate, and the needed bond dimension depends on the entanglement generated by measuring the noise-corrupted state.
The measurement basis does not affect syndrome measurement outcomes, and any basis, even a non-product basis within the row, can be chosen.
The choice can, however, affect the entanglement and bond dimension of the MPS.
Here, we measure all qubits in the $X$ basis.

\section{Additional numerics data}
In this section, we provide additional numerical data for the ML decoding in the surface code under two unitary error models: (1) single and two-qubit Pauli-X rotations (Sec.~\ref{sec:num_coh-X_coh-XX}).
; (2) general single-qubit rotation and two-qubit Pauli-X rotation (Sec.~\ref{sec:num_coh-XYZ_coh-XX}).

\subsection{X rotation and XX rotation}\label{sec:num_coh-X_coh-XX}
Here, we provide additional numerical results for ML decoding in the surface code with single- and two-qubit Pauli-X rotation. 

\begin{figure}
\centering
\begin{tikzpicture}[scale=0.7]
\node[scale=0.85] at (-8,0) {\includegraphics[width=2.5in]{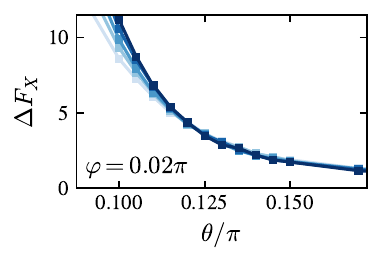}};
\node[scale=0.85] at (-8,-5) {\includegraphics[width=2.5in]{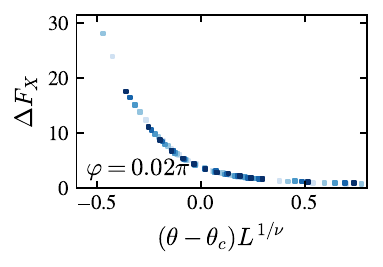}};
\node[scale=0.85] at (0,0) {\includegraphics[width=2.5in]{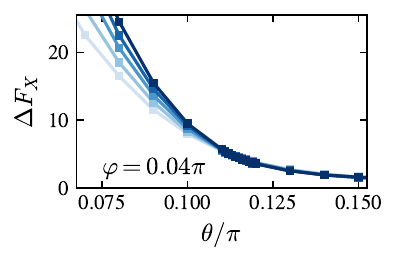}};
\node[scale=0.85] at (0,-5) {\includegraphics[width=2.5in]{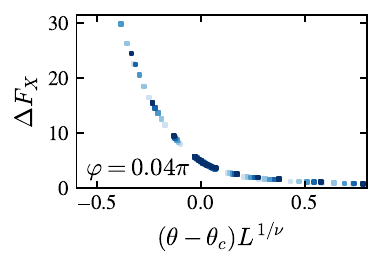}};
\node[scale=0.85] at (8,0) {\includegraphics[width=2.5in]{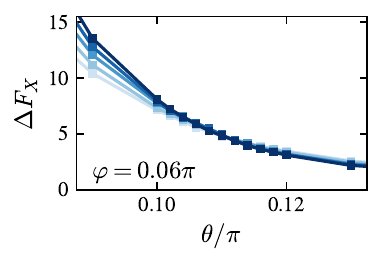}};
\node[scale=0.85] at (8,-5) {\includegraphics[width=2.5in]{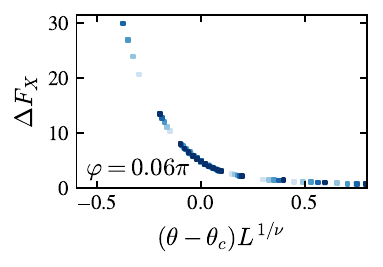}};
\node[scale=0.85] at (-8,-10) {\includegraphics[width=2.5in]{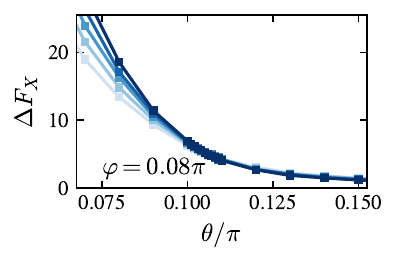}};
\node[scale=0.85] at (-8,-15) {\includegraphics[width=2.5in]{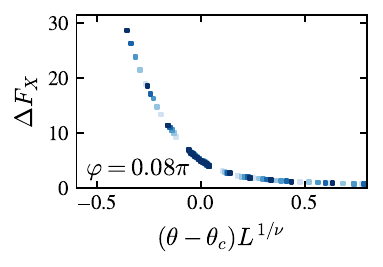}};
\node[scale=0.85] at (0,-10) {\includegraphics[width=2.5in]{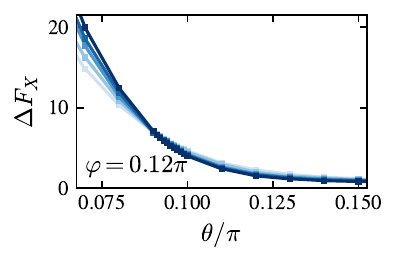}};
\node[scale=0.85] at (0,-15) {\includegraphics[width=2.5in]{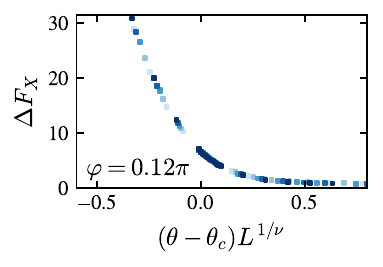}};
\node[scale=0.85] at (8,-10) {\includegraphics[width=2.5in]{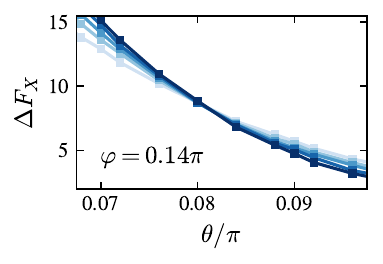}};
\node[scale=0.85] at (8,-15) {\includegraphics[width=2.5in]{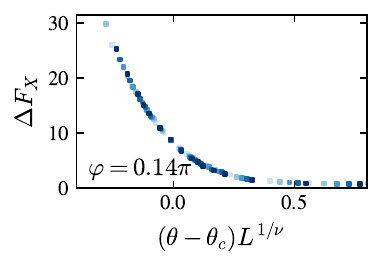}};
\node[scale=0.85] at (-8,-20) {\includegraphics[width=2.5in]{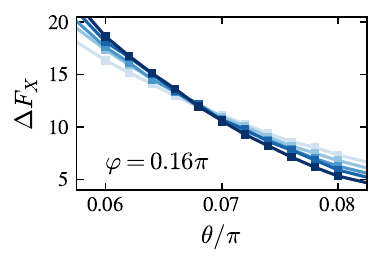}};
\node[scale=0.85] at (-8,-25) {\includegraphics[width=2.5in]{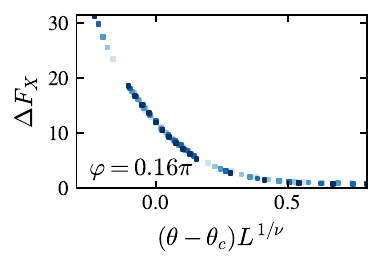}};
\node[scale=0.85] at (0,-20) {\includegraphics[width=2.5in]{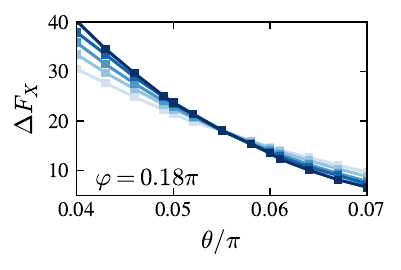}};
\node[scale=0.85] at (0,-25) {\includegraphics[width=2.5in]{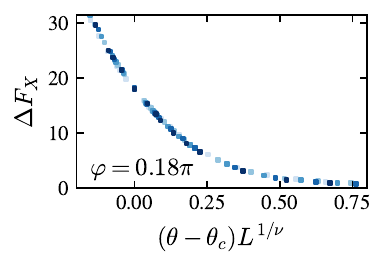}};
\node[scale=0.85] at (8,-20) {\includegraphics[width=2.5in]{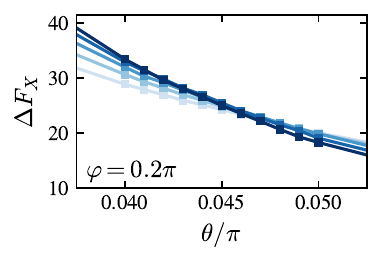}};
\node[scale=0.85] at (8,-25) {\includegraphics[width=2.5in]{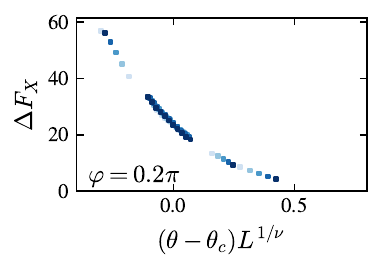}};
\end{tikzpicture}
\caption{Defect free energy as a function of $\theta$ for fixed $\phi = 0.02\pi, 0.04\pi, 0.06\pi, 0.08\pi, 0.12\pi, 0.14\pi, 0.16\pi, 0.18\pi, 0.2\pi$. Data collapse are obtained using the ansatz $\Delta F_X = f((\theta - \theta_c)L^{1/\nu})$. The results are obtained in the simulation with aspect ratio $L_y/L_x = 4$ and bond dimension $\chi = 128$.}
\label{suppfig:coh-X_coh-XX}
\end{figure}

In Fig.~\ref{suppfig:coh-X_coh-XX}, we present the defect free energy as a function of $\theta$ for various fixed $\varphi$ and obtain finite-size scaling collapse using the ansatz $\Delta F_X = f( ( \theta - \theta_c) L^{1/\nu})$. The extracted critical rotation angles $(\theta_c, \varphi_c)$ and correlation length exponents $\nu$ are summarized in Tab.~\ref{tab:critical_points} and presented in Fig.~1(c, d) of the main text.

\begin{table}[]
    \centering
    \begin{tabular}{c|c|c|c|c|c|c|c|c|c|c|c}
    \hline\hline
    $\varphi_c/\pi$ & 0.02 & 0.04 & 0.06 & 0.08 & 0.102 & 0.12 & 0.14 & 0.16 & 0.18 & 0.2 & 0.22 \\
    $\theta_c/\pi$ &  0.123 & 0.113& 0.11& 0.106& 0.102& 0.091& 0.081& 0.068& 0.055& 0.046& 0.040 \\
    $\nu$     & 2.215 & 2.368& 2.409& 2.416& 2.270 & 2.258& 2.186& 1.965& 1.779& 1.607& 1.541 \\
    \hline\hline
    \end{tabular}
    \caption{Critical rotation angles $(\theta_c,\varphi_c)$ and correlation-length critical exponents $\nu$ at the ferromagnetic transition in Fig~1(c) of the main text.}
    \label{tab:critical_points}
\end{table}

\begin{figure}
    \centering
    \begin{tikzpicture}
        \node[scale=0.85] at (0,0) {\includegraphics[width=3.4in]{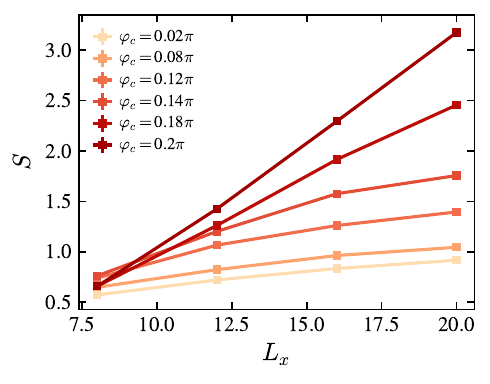}};
        \node[scale=0.85] at (8,0) {\includegraphics[width=3.4in]{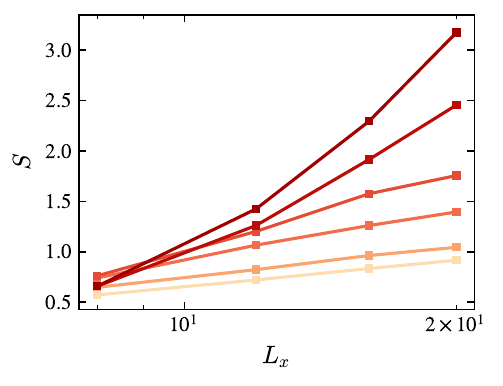}};
        \node at (-3.5,3) {(a)};
        \node at (4.5,3) {(b)};
    \end{tikzpicture}
    \caption{Half-chain entanglement entropy in the steady state of transfer matrix contraction at the critical point $(\theta_c,\varphi_c)$ (in Tab.~\ref{tab:critical_points}) as a function of system size $L_x$ in log-linear plot (panel a) and log-log plot (panel b). The entanglement scales logarithmically in $L_x$ for $\varphi_c \lesssim 0.1\pi$ and scales linearly for $\varphi_c \gtrsim 0.1\pi$. The simulation is performed with bond dimension up to $128$, and the results are averaged over up to $1000$ samples.}
    \label{suppfig:coh-X_coh-XX_critical_entanglement}
\end{figure}

In Fig.~\ref{suppfig:coh-X_coh-XX_critical_entanglement}, we present the scaling of half-chain entanglement entropy at the critical point $(\theta_c, \varphi_c)$ in Tab.~\ref{tab:critical_points}. 
We observe that the entanglement entropy scales logarithmically in the system size for $\varphi_c \lesssim 0.1\pi$ and exhibits a volume-law scaling for $\varphi_c \gtrsim 0.1\pi$. This supports the presence of a volume-law ferromagnetic phase (in Fig.~1(a) of the main text).

\subsection{X, Y, Z rotation and XX rotation}\label{sec:num_coh-XYZ_coh-XX}
We consider the unitary error that consists of single-qubit rotations $e^{\ri\theta_x Z_\ell + \ri \theta_y Y_\ell}$ and two-qubit rotations $e^{\ri \varphi X_\ell X_{\ell'}}$.
We demonstrate the existence of a ferromagnetic volume -law (FM VL) phase.

\begin{figure}
    \centering
    \includegraphics[width=3.4in]{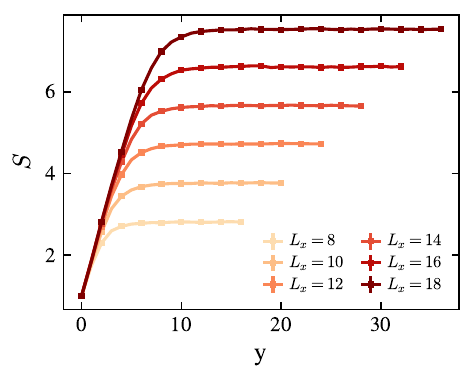}
    \caption{Entanglement entropy as a function of contraction step $y$ for $(\theta_x, \theta_y) = (0.2 \pi, 0.06\pi)$, and $\varphi = 0.2\pi$. 
    }
    \label{suppfig:coh-XYZ_coh-XX_entanglement_time}
\end{figure}

To start, we take $\theta_y = 0$ and $(\theta_x, \varphi) = (0.2\pi, 0.2\pi)$, which is deep in the volume-law paramagnetic phase [in Fig.~1(a) of the main text].
The unitary error with a non-zero $\theta_y$ creates both $e$ and $m$ anyons in the surface code.
In this case, the optimal decoding for $X$ and $Z$ errors is no longer independent, and the stat-mech model involves two copies of the Ising spins.
The Ising spins for the Z errors exhibit an FM order when $\theta_y$ is below a threshold $\theta_c$, indicating a phase that encodes the classical information.
Increasing $\theta_y$ leads to an FM transition within the volume-law phase.
We first present additional numerical data for half-chain entanglement entropy as a function of contraction step $y$ in Fig.~\ref{suppfig:coh-XYZ_coh-XX_entanglement_time}; the steady state entanglement of this dynamics is presented in Fig.~3(b) of the main text.
For $\theta_y = 0.06 \pi$, the entanglement grows and saturates to a value linear in the system size, indicating a volume-law entanglement.

\begin{figure}
    \centering
    \begin{tikzpicture}
        \node[scale=0.85] at (0,0) {\includegraphics[width=3.4in]{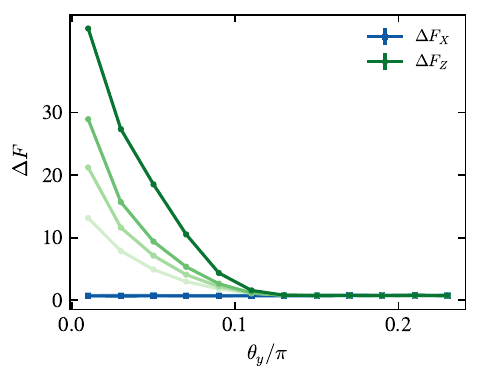}};
        \node[scale=0.85] at (8,0) {\includegraphics[width=3.4in]{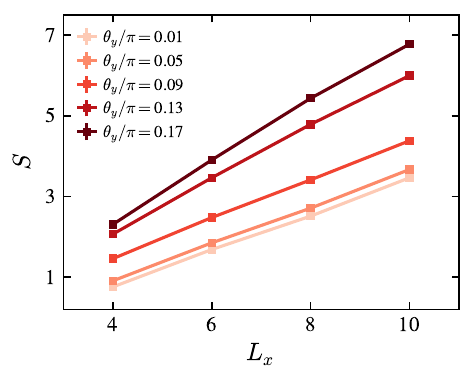}};
        \node at (-3.5,3) {(a)};
        \node at (4.5,3) {(b)};
    \end{tikzpicture}
    \caption{Decoding transition in the surface code with unitary error $e^{\ri\theta_x X_\ell + \ri \theta_y Y_\ell}$ and $e^{\ri \varphi X_{\ell}X_{\ell'}}$. The parameters $\theta_x$ and $\varphi$ are set to a fixed value, $(\theta_x, \varphi) = (0.2\pi,0.2\pi)$. (a) Defect free energies $\Delta F_Z$ and $\Delta F_X$ as a function of $\theta_y$ for various system sizes $L_x = 4,6,8,10$, and aspect ratio $L_y/L_x = 2$. Curves with increasing opacity represent an increasing system size. (b) Half-chain entanglement entropy in the steady state as a function of $L_x$ for various $\theta_y$. The simulation is perform with bond dimension up to $256$. The results are averaged up to $1000$ samples, and the statistical errors are invisible on the plot.}
    \label{fig:coh-XYZ_coh-XX_thetay}
\end{figure}

Next, we present the numerical results about the transition within the volume-law phase when increasing $\theta_y$.
The transition is detected by the free energy cost $\Delta F_Z$ of changing the boundary condition in the spins associated with Pauli-Z strings.
Due to the large entanglement during the contraction, we can only estimate a threshold $\theta_{y,c} \sim 0.13\pi$.
The defect free energy $\Delta F_Z$ is linear in its size below the threshold, indicating the encoding of classical information; it becomes a constant above the threshold, as shown in Fig.~\ref{fig:coh-XYZ_coh-XX_thetay}(a).
The defect free energy $\Delta F_X$ associated with the boundary condition change of the spins associated with Pauli-X strings is always a constant, indicating the loss of classical information [Fig.~\ref{fig:coh-XYZ_coh-XX_thetay}(a)].
The half-chain entanglement entropy remains volume-law when tuning $\theta_y$ across the transition.

\bibliography{refs}